\documentclass[twocolumn]{aastex631}

\usepackage{amsmath}
\usepackage{mathtools}
\usepackage{tabularx}
\usepackage{url}

\usepackage{soul}  

\shorttitle{Exoplanet Plasma Tori}
\shortauthors{Boehm et al.}

\begin{document}

\title{Constraining Ongoing Volcanic Outgassing Rates and Interior Compositions of Extrasolar Planets with Mass Measurements of Plasma Tori}

\shortauthors{Boehm et al.}

\author[0000-0003-3987-3776]{V. Abby Boehm}
\affiliation{Department of Astronomy and Carl Sagan Institute, Cornell University, 122 Sciences Drive, Ithaca, NY, 14853, USA}

\correspondingauthor{V. Abby Boehm}
\email{vab55@cornell.edu}
\author[0000-0002-0726-6480]{Darryl Z. Seligman}
\altaffiliation{NSF Astronomy and Astrophysics Postdoctoral Fellow}
\affiliation{Department of Physics and Astronomy, Michigan State University, East Lansing, MI 48824, USA}

\author[0000-0002-8507-1304]{Nikole K. Lewis}
\affiliation{Department of Astronomy and Carl Sagan Institute, Cornell University, 122 Sciences Drive, Ithaca, NY, 14853, USA}

\begin{abstract}
We present a novel method of constraining volcanic activity on extrasolar terrestrial worlds via characterization of circumstellar plasma tori. Our work generalizes the physics of the Io plasma torus to propose a hypothetical circumstellar plasma torus generated by exoplanetary volcanism. The quasi-steady torus mass is determined by a balance between material injection and ejection rates from volcanic activity and corotating magnetospheric convection, respectively. By estimating the Alfv\'en surfaces of planet-hosting stars, we calculate the torus mass-removal timescale for a number of exoplanets with properties amenable to plasma torus construction. Assuming a uniform toroidal geometry comparable to Io's ``warm'' torus, we calculate quasi-steady torus masses inferable from the optical depth of atomic spectral features in torus-contaminated stellar spectra. The calculated quasi-steady masses can be used to constrain the volcanic outgassing rates of each species detected in the torus, providing quantitative estimates of bulk volcanic activity and interior composition with minimal assumptions. Such insight into the interior state of an exoplanet is otherwise accessible only after destruction via tidal forces. We demonstrate the feasibility of our method by showcasing known exoplanets which are susceptible to tidal heating and could generate readily-detectable tori with realistic outgassing rates of order 1~ton~s$^{-1}$, comparable to the Io plasma torus mass injection rate. This methodology may be applied to stellar spectra measured with ultraviolet instruments with sufficient resolution to detect atomic lines and sensitivity to recover the ultraviolet continuum of GKM dwarf stars. This further motivates the need for ultraviolet instrumentation above Earth's atmosphere.
\end{abstract}


\section{Introduction} \label{sec:intro}
Volcanic activity within and beyond the Solar System provides crucial insight into planetary formation, atmospheric composition, and habitability. Volcanic activity within the first $\sim1\textrm{ Gyr}$ post-formation is thought to be a primary mechanism by which terrestrial planets build secondary atmospheres \citep[e.g.,][]{Kasting1993Sci...259..920K} and is therefore critical in explaining secondary atmosphere masses and compositions. Furthermore, observing volcanic activity provides insight into the mantle volatile content of terrestrial planets, e.g. by revealing the abundance of sulfur compounds and deficit of carbon compounds in the interior of Io \citep{Kesz2023ASSL..468..211K}. This information is typically inaccessible until an exoplanet has been destroyed by tidal disruption \citep[e.g.][]{Duvvuri_2020}. Terrestrial exoplanets on short-period, eccentric orbits should be subject to significant tidal heating from their host stars \citep{Goldreich1963,Hut1981,Driscoll2015,Becker2023,Becker2024}. This heating should sustain long-term and active surface volcanism, analogous to the tidal heating in the Jupiter-Io system \citep[e.g.,][]{Jackson2008ApJ...681.1631J, Henning2009ApJ...707.1000H, Seligman2024ApJ...961...22S}. Tidally-heated, volcanically-active exoplanets can therefore provide valuable windows into volcanic outgassing and internal composition, an understanding of which is necessary in order to trace the formation, evolution, and potential habitability of terrestrial worlds.

To date, confirmation of previous or ongoing volcanic activity in the Solar System has largely relied on direct surface imaging of geologic features consistent with volcanism. Examples of these features include frozen basalt deposits imaged in lunar mare \citep[e.g,][]{Head1992GeCoA..56.2155H}, dormant shield volcanoes imaged on Mars \citep{McCauley1972Icar...17..289M}, active lava lakes and volcanic plumes revealed in \textit{Voyager} images of Io \citep{Smith1979,Morabito1979Sci...204..972M,Strom1979Natur.280..733S}, and volcanoes on Venus \citep{Barsukov1986JGR....91D.378B} which were confirmed to have been active during the \textit{Magellan} mission \citep{herrickhensley}.

Unlike volcanism within our Solar System, inferences of past or ongoing exoplanetary volcanism require indirect methods without resolved surface images. For example, transmission spectroscopy has permitted the tentative detection of molecules such as CO$_2$ and SO$_2$ in the atmospheres of rocky worlds \citep[e.g.,][]{Hu55Cnc2024Natur.630..609H,GressierL98b2024ApJ...975L..10G,Belloarufe2025}. These tentative detections have been interpreted to suggest the occurrence of volcanic activity within the past 100s to 1000s of Myrs. However, CO$_2$ can persist over many Gyr even in the presence of carbon-cycling processes such as plate tectonics, ocean sequestration, and respiration by living organisms \citep{WongCO2}. The presence of CO$_2$ therefore does not require that volcanism be presently active. Alternatively, the presence of SO$_2$ would be a robust indicator of active volcanism due to its short photochemical lifetime and lack of alternative formation pathways \citep{sulfurtextbook}. However, the rapid photodissociation timescale of SO$_2$ ---particularly in the upper atmosphere where it is most amenable to detection via transmission spectroscopy--- may prohibit its accumulation to robustly measurable levels \citep{KaltHen2010AJ....140.1370K}. Eclipse spectroscopy of rocky worlds can provide an unambiguous indication of volcanic activity by revealing surface composition, for example via the detection of volcanic rocks such as basalt \citep{First2024NatAs.tmp..275F}. However, as basalt is a stable and long-lived material, its presence alone cannot be used to differentiate between active and dormant volcanism. In summary, current methods can detect signatures of exoplanetary volcanic activity \textit{at some point}, but proving that the activity is ongoing requires a new approach.

The Jupiter-Io system offers a promising path forward for remote sensing of active volcanism in exoplanetary systems. Direct images obtained by the \textit{Voyager 1} flyby confirmed Io's ongoing volcanism \citep{Smith1979, Morabito1979Sci...204..972M, Strom1979Natur.280..733S} shortly after \citet{Peale1979Sci...203..892P} predicted substantial surface volcanism based on the theory of tidal heating developed in \citet{Peale1978Icar...36..245P}. Radio and optical signatures of Io's volcanic activity had  been detected prior to the flyby. However, these signals had not been correctly attributed to outgassing. Specifically, Io's orbital motions were stimulating decametric emissions from Jupiter's poles \citep{Bigg1964Natur.203.1008B}, implying that the Jovian magnetosphere contained plasma of unknown and possibly Ionian origin \citep{Siscoe1977Icar...31....1S}. Post-eclipse brightening of Io's surface was interpreted as evidence of a tenuous, collapsible atmosphere which freezes out onto the surface as a reflective frost whenever Io is in Jupiter's shadow \citep{Binder1964Icar....3..299B}. While those observations proved difficult to reproduce, subsequent studies and models also concluded that Io had a tenuous atmosphere \citep{McElroy1975ApJ...196..227M,Fink1976Icar...27..439F,Cruikshank1978ApJ...225L..89C}. Detections of toroidally-distributed neutral sodium emission surrounding the orbit of Io \citep{Brown1974ApJ...187L.125B, Trafton1974ApJ...190L..85T,Bergstralh1975ApJ...195L.131B} implied that the tenuous atmosphere was undergoing escape. Both the magnetospheric plasma and the escaping atmosphere pointed to ongoing volcanism on Io. However, correct interpretation of these lines of evidence required additional context from \textit{Voyager 1} observations.

While \textit{Voyager 1} images confirmed the presence of volcanic activity on Io's surface, measurements from \textit{Voyager 1}'s plasma detector \citep{Bridge1977SSRv...21..259B} and ultraviolet (UV) spectrometer \citep{Broadfoot1979Sci...204..979B} revealed a torus of plasma spanning between 5 to 40 R$_{\rm J}$ corotating with the Jovian magnetosphere. Energy/charge spectra and emission spectroscopy indicated a composition of sulfur and oxygen ions \citep{Bridge1979Sci...204..987B, Broadfoot1979Sci...204..979B}. Later analyses refined the distribution of the torus such that 2~Mt or $\sim90\%$ of its mass was concentrated at and around Io's orbit at 6 R$_{\rm J}$ \citep{BagenalDISTRO1980GeoRL...7...41B}. Plasma density within the torus averages between 1000 to 3000 electrons per cm$^3$. This number density decays rapidly in the direction of Jupiter, and  more gradually in the direction of Europa \citep{Shemansky1980ApJ...236.1043S, Strobel1980ApJ...238L..49S}. Io's orbital motion through the plasma torus generates a wake of Alfv\'en waves in the magnetic field lines that connect Io and Jupiter's poles. These Alfv\'en waves induce high-energy electron currents onto the Jovian poles which in turn drive aurorae and decametric emissions \citep{Deift1973P&SS...21.1417D, Thorne1979GeoRL...6..649T, Goertz1980JGR....85.2949G}. Infrared (IR) spectra collected by \textit{Voyager 1} also confirmed the existence of Io's thin atmosphere, composed of volcanically-supplied SO$_2$ with trace alkali elements \citep{Pearl1979Natur.280..755P, Kumar1979Natur.280..758K}.

Io's atmosphere and plasma torus are remarkably fragile structures that are only sustained via volcanic activity. Constant bombardment of Io's atmosphere by the plasma torus drives thermodynamic escape \citep{Pospie1992GeoRL..19..949P}. This bombardment in concert with solar-driven erosion would completely remove Io's atmosphere if not for frequent replenishment by volcanic eruptions. Likewise, the plasma torus itself also requires replenishment via Io's volcanism. The non-Keplerian plasma velocity arises from corotation with the Jovian magnetosphere forced by magnetic coupling\citep{Alfven1943ArMAF..29B...1A}. Corotating plasma experiences outwards radial drift via magnetospheric convection on a timescale of $\sim10-100$ days \citep{Hill1981} until the Jovian magnetic field can no longer retain it against the solar wind pressure \citep{ShemanskyDIFFUSION1980ApJ...242.1266S}. These processes drive torus plasma removal at a rate of approximately 1~ton~s$^{-1}$. This requires an equivalent torus mass injection rate to sustain the torus \citep{Dessler1980Icar...44..291D}, although there is empirical evidence for small-scale variability of this rate \citep{Richardson1980GeoRL...7...37R}. The existence of the torus despite the rapid dispersal timescale uniquely and unambiguously provides evidence for both the presence and recency of volcanic activity on Io. Furthermore, the torus can be observed remotely, enabling characterization of Io's active volcanism independent of resolved imaging.

Although the Io plasma torus is the only known example of such a structure, the underlying physical processes are not unique to the Jupiter-Io system and can be reasonably extended to star-planet systems. Plasma torus formation generically requires an outgassing rocky body prone to atmospheric escape orbiting within the magnetosphere of a rotating object. Main-sequence stars are typically magnetically active with surface field strengths ranging from $\sim100$~G to upwards of 1000~G for the most active M-dwarf stars \citep{Moutou2017MNRAS.472.4563M,CARMENES2019A&A...626A..86S}. Magnetic confinement of plasma within a stellar magnetosphere has been observed around ultracool stars \citep{Kao2023Natur.619..272K} and could hypothetically be possible in the magnetospheres of larger stars, provided the magnetic field confinement can outweigh the effects of stellar activity and radiation. More than 2000 of the currently-known exoplanets orbit closer to their stars than Mercury orbits the Sun \citep{ARCHIVE2020ipac.data..N12N}, and therefore likely within the stellar magnetosphere. Approximately 400 of these short-period planets have radii consistent with terrestrial (i.e., $R\leq1.6$~R$_{\rm \oplus}$ \citep[][]{Valencia2006, Valencia2007, Dorn2015, Fulton2017}, of which $\sim70$ are tentatively to confidently eccentric, and thus could be volcanically active through tidal heating. Therefore, the conditions to build circumstellar plasma tori may exist in several known planetary systems. The detection of a circumstellar plasma torus in conjunction with a short-period rocky exoplanet would be strong evidence for active and ongoing exoplanetary volcanic activity.

Here we describe a novel approach to detect and constrain volcanic activity on exoplanets via circumstellar plasma tori. In Section \ref{sec:methods} we describe the physical processes driving the formation, retention, and dispersal of plasma tori. We present a relationship between the physical conditions of the star-planet system and the volume and minimum steady state mass of a hypothetical plasma torus. Prior studies have posited that a plasma torus in a transiting exoplanetary system would be detectable by UV spectroscopy if the column density of the torus on the line of sight was of order $10^{12}-10^{13}$~cm$^{-2}$ \citep{Kislya2018ApJ...858..105K, Kislya2019arXiv190705088K}. In Section \ref{sec:analysis}, we expand upon this work and show how the strength of these absorption signatures (i.e. the column density of the torus) can be combined with an approximated torus geometry and the derived minimum steady-state torus mass to estimate the volcanic outgassing rate of the source planet. In Section \ref{sec:results}, we present the most promising candidates for detection of volcanically-outgassed plasma tori. We demonstrate that absorption signatures readily detectable by the Hubble Space Telescope (HST) could be produced from volcanic outgassing rates comparable to those observed in our Solar System, provided that the UV flux of identified targets overcomes noise floor limitations. In Section \ref{sec:caveats} we discuss caveats of our method and feasibility for current instrumentation. Where the noise floor becomes an obstacle, we instead highlight proposed and forthcoming UV-sensitive observatories for which this methodology may be accessible. Thus, our work further underscores the need for space-based UV-sensitive missions beyond HST to characterize the atmospheres and exospheres of exoplanets. In Section \ref{sec:conclusions} we summarize our work and provide recommendations for future studies.

\section{The Physics of a Planetary System-Scale Plasma Torus} \label{sec:methods}

In this section we scale the Jupiter-Io plasma torus to a star-planet system. We consider a rocky planet that is volcanically active and outgassing volatiles at a rate $\dot M_{\rm volc}$. The planet is orbiting within the magnetosphere of a magnetically-active star with semimajor axis $a$, eccentricity $e$, and orbit period $P_{\rm Orb}$. The star has mass $M_*$, radius $R_*$, effective temperature $T_{\rm eff}$, spin rate $\Omega$, and surface magnetic field strength $B_*$. We assume that (i) the stellar wind is driving mass loss from the star at a rate $\dot M_{\rm wind}$, (ii) the stellar spin rate and stellar magnetosphere spin rate are the same, and (iii) the stellar spin and planetary orbit angular momentum vectors are aligned.

Here we summarize the physics driving plasma torus formation. We assume that some fraction $f_{\rm esc}\leq1$ of the outgassed material escapes, becomes ionized, and forms a rarefied and highly conductive plasma. In the presence of the stellar magnetic field the plasma acts as an ideal magnetohydrodynamic (MHD) fluid which is confined to the field lines \citep[for a full derivation, see Chapter 2 of][]{EricPriest2014masu.book.....P}. The plasma is ``picked up'' and carried away along the orbit path at the magnetospheric corotation rate $\Omega$ by the rotating stellar magnetic field. Provided that the stellar rotation rate is not equal to the Keplerian orbit frequency $\Omega_{\rm Kep}=2\pi/P_{\rm Orb}$, the magnetically-confined ions will disperse along the orbit path and form a diffuse torus of plasma. Note that in some cases there exists commensurability between the stellar spin and orbital periods (see Table 1 in \citet{Lanza2022}) which could impede dispersal of the plasma; we do not consider such cases here. In Sections \ref{sec:supply}, \ref{sec:confine}, and \ref{sec:radiation} we provide greater detail about the injection and confinement of material within the torus.

This plasma torus is a remarkably transient structure because its dispersal operates on a magnetospheric convection timescale $\tau$ that is brief relative to geologic timescales. The plasma is removed via corotating magnetospheric convection, a process outlined by \citet{Hill1981}. In this process, over-densities in the inner plasma torus generate currents that create radially-outwards electric fields. These electric fields act as a centrifugal force, driving magnetic flux tubes that contain overly dense plasma to ``sink'' radially outwards. The magnetic confinement force operating on the over-dense flux tubes decreases until it cannot maintain corotation and the plasma in the tube escapes. The now under-dense flux tubes ``float'' radially inwards until they return to the source planet, which reloads them with fresh plasma and the cycle repeats. A sustained torus requires continuous injection of material on a timescale comparable to the magnetospheric convection timescale which is of order weeks to months (Section \ref{sec:results}). Therefore, the detection of a plasma torus implies the source planet underwent \textit{recent and ongoing} mass loss. In Sections \ref{sec:timescale}, \ref{sec:span}, and \ref{sec:steadystate} we provide an overview of these processes and calculate the relationship between convective timescale, torus mass, stellar parameters (e.g. mass, radius, magnetic field strength, spin rate), and planet orbit configuration.

\subsection{Supplying Material to the Torus} \label{sec:supply}
A plasma torus will efficiently sustain atmospheric escape via collisional heating \citep{Pospie1992GeoRL..19..949P}, regardless of the mechanism that initiated mass loss. We assume that a substantial amount of atmospheric escape is already occurring as a result of photoevaporation (see Appendix A for details). This escaping material diffuses out into the exosphere of the volcanic exoplanet, creating a structure analogous to Io's ``neutral cloud'' \citep[see e.g. Section 4.1 of][]{BagenalDols2020}. Due to stellar X-ray and extreme UV (XUV) radiation, some fraction of the neutral cloud material is ionized and subsequently entrained in the stellar magnetic field. The now-corotating plasma is accelerated by the stellar magnetic field to travel at a rate not equal to the Keplerian orbit rate $\Omega_{\rm Kep}$. In the process of being accelerated by the stellar magnetic force, the ions acquire a pickup energy of $E_{\rm pickup}=\frac{1}{2}m\Delta v^2$, where $\Delta v$ is the difference between the corotation velocity and the Keplerian orbit velocity. The ions then impart that energy back into the atmosphere of the planet through collisions. For exoplanets with orbital periods $P_{\rm Orb}\lesssim10$~days that are not commensurate with their host star spin periods, pickup of lost plasma can impart 10s to 100s of eV into each ion. As $E_{\rm pickup}$ will typically exceed even the highest first ionization energy -- that of helium at $E_{\rm ionization}=25$~eV -- collisions between corotating plasma and atmospheric atoms will be energetic enough to dissociate molecules and cause escape and ionization of the neutral atoms within the planetary atmosphere and neutral cloud. Thus, corotation effectively transfers rotational energy from the star into thermal energy imparted by collisions of the corotating plasma with the planet as a feedback loop. Plasma heating can produce two equilibrium states:
\begin{itemize}
    \item All of the planet's atmospheric material escapes ($f_{\rm esc}\rightarrow1$) if no other mechanism inhibits the heating. In this case the planet will be airless while the plasma heating reaches a maximum rate set by the volcanic activity of the planet. The torus mass injection rate, in turn, will equal the full extant of the planetary volcanism.
    \item Heating generates current loops in the planet's ionosphere which, if sufficiently strong, will deflect some of the incoming plasma \citep{Johnson1993GeoRL..20.1735J}. This prevents the plasma heating from reaching its maximum potential level ($f_{\rm esc}<1$). The torus mass injection rate is less than the planetary volcanic outgassing rate.
\end{itemize}
The latter scenario results in saturation at a quasi-steady equilibrium between plasma heating and atmospheric escape. This is what permits Io to retain a thin atmosphere with surface pressure $p_{\rm surf}\sim10^{-9}$~bar rather than undergoing complete atmosphere loss \citep{YungDemore1999ppa..conf.....Y}.

Note that while establishing ionospheric currents can deflect plasma heating, this mechanism provides no defense against photoevaporation driven by stellar radiation. For exoplanets orbiting within the stellar magnetosphere, substantial atmospheric loss driven by stellar radiation (see Appendix \ref{sec:atmosescape} for details) may cause significant erosion of the ionosphere and thus limit the strength of plasma-deflecting ionospheric currents. As a result, the quasi-steady equilibrium discussed above may be inaccessible to such exoplanets. We therefore expect the majority of these worlds to be found in the $f_{\rm esc}\simeq1$ maximum plasma heating state and as a consequence to be bereft of substantial atmospheres.

In both cases, plasma heating drives atmosphere loss and ionization of escaping material once the torus is established. The torus will be supplied with mass as long as the planet is volcanically active.

\subsection{The Conditions for Magnetic Confinement} \label{sec:confine}
Io's torus plasma is retained only where the magnetic confinement force exerted by Jupiter's magnetosphere is stronger than the solar wind pressure. Likewise, circumstellar plasma can only be retained where the stellar magnetic confinement force dominates over the stellar wind pressure. The wind magnetic confinement parameter $\eta(r)$ is defined as the ratio of the magnetic field energy density to the kinetic energy density of the stellar wind as a function of distance $r$ from the star \citep{ud2002ApJ...576..413U}. The magnetic confinement force dominates and a plasma torus can form when $\eta(r)$ is greater than 1.

We assume that the plasma escapes the torus where the magnetic and wind energy density are equal ($\eta=1$), typically referred to as the Alfv\'en radius $R_A$. The Solar Alfv\'en radius is estimated to reside between 10 to 60 solar radii and exhibits significant spatial and temporal irregularity \citep[see e.g.,][]{Goelzer2014JGRA..119..115G, Deforest2014ApJ...787..124D, Liu2021ApJ...908L..41L, Cranmer2023SoPh..298..126C}. For the order of magnitude calculation in this paper, we estimate the Alfv\'en radius using the method presented by \cite{Owocki2009EAS....39..223O}. In this method, we assume that the stellar magnetosphere is azimuthally symmetric and dipolar in form:
\begin{equation}\label{eqn:bfield}
    B(r)=B_*\bigg(\frac{r}{R_*}\bigg)^{-3}\,.
\end{equation}
We parameterize the stellar wind velocity $v$ as a function of distance $r$:
\begin{equation} \label{eqn:vinf}
    v(r)=v_\infty\bigg(1-\frac{R_*}{r}\bigg)^b\,.
\end{equation}
In Equation \ref{eqn:vinf}, $v_\infty$ is the terminal stellar wind velocity and $b$ is a power index typically assumed to be 1. $v_\infty$ is often comparable to and therefore approximated by the stellar surface escape velocity, $v_\infty\sim\sqrt{2GM_*/R_*}$. Under these assumptions, the wind magnetic confinement parameter is given by
\begin{equation} \label{eqn:eta}
    \eta(r)=\eta_* \bigg[\bigg(\frac{r}{R_*}\bigg)^{4}-\bigg(\frac{r}{R_*}\bigg)^{3}\bigg]^{-1}\,.
\end{equation}
In Equation \ref{eqn:eta}, $\eta_*$ is the wind magnetic confinement parameter at the stellar surface:
\begin{equation} \label{eqn:etastar}
    \eta_*=\underbrace{\bigg(\frac{B_*^2}{2\mu_0}\bigg)}_\textrm{Magnetic Field}\,\underbrace{\bigg(\frac{\dot M_{\rm wind}v_\infty}{R_*^2}\bigg)^{-1}}_\textrm{Kinetic Wind}\,.
\end{equation}
In Equation \ref{eqn:etastar}, $\mu_0$ is the magnetic permeability in vacuum. The solution to Equation \ref{eqn:eta} for $\eta(R_A)=1$ yields the Alfv\'en radius scaled by the stellar radius,
\begin{equation} \label{eqn:AlfvenR}
  \bigg(\frac{R_A}{R_*}\bigg)^{4}-\bigg(\frac{R_A}{R_*}\bigg)^{3}=\eta_*\,.
\end{equation}

\begin{figure}[h!]
    \centering
    \includegraphics[width=1.00\linewidth]{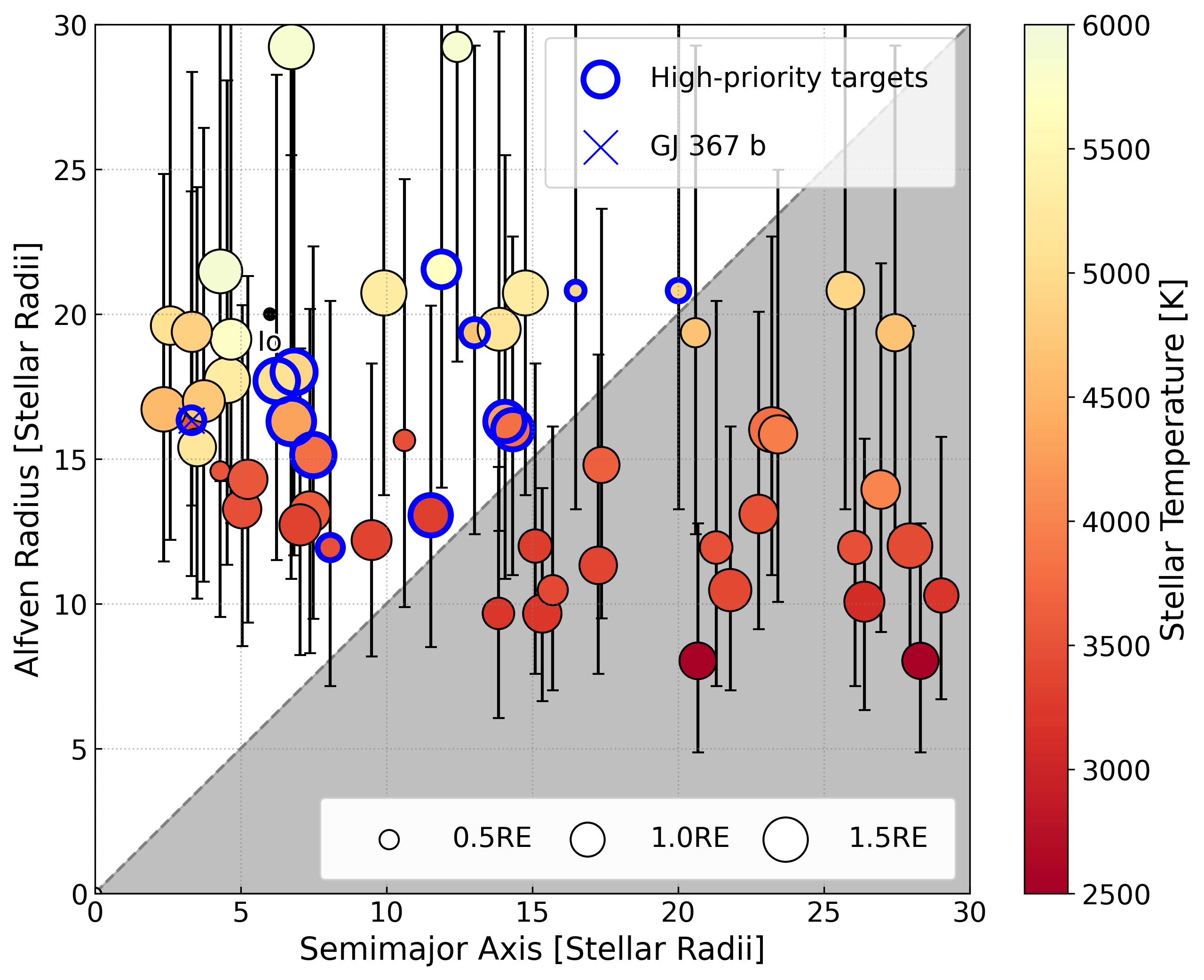}
    \caption{Stellar Alfv\'en radii of stars with measured masses, radii, and spin periods which are also known to host terrestrial exoplanets. For the 33 exoplanets which orbit at $a<R_A$, magnetic confinement is strong enough to permit construction of plasma tori. Io is plotted in terms of Jupiter radii R$_{\rm J}$; the Jovian Alfv\'en radius is sourced from \cite{McNutt1979Natur.280..803M} and \cite{Hill1980Sci...207..301H}. Targets in the grey region orbit outside of the stellar Alfv\'en radius. Conditions for being classified as a high-priority target, including the classification of GJ~367~b as highest priority, are described in Section \ref{sec:results}.}
    \label{fig:alfvenradius}
\end{figure}

In Figure \ref{fig:alfvenradius}, we compare the stellar Alfv\'en radius computed with Equation \ref{eqn:AlfvenR} against the orbit semimajor axes of terrestrial exoplanets. We define ``terrestrial'' planets as those with radius $R\leq1.6$~R$_{\oplus}$  \citep[][]{Valencia2006, Valencia2007, Dorn2015, Fulton2017}. Alfv\'en radius computation requires some estimation of the stellar $B_*$ and $\dot M_{\rm wind}$. For GKM stars of all ages and rotational periods, $B_*$ and $\dot M_{\rm wind}$ can take values spanning 10--1000~G and $10^{-15}$--$10^{-10}$~M$_\odot$~yr$^{-1}$ respectively \citep{Moutou2017MNRAS.472.4563M,CARMENES2019A&A...626A..86S, Bloot2025arXiv250214701B}. Stellar winds and magnetic fields in GKM dwarf stars are closely tied to age and stellar rotation period $P_*$, with greater age and slower spins correlating with weaker magnetic fields and lower wind mass loss rates \citep[e.g.][]{Wright2018MNRAS.479.2351W,CARMENES2019A&A...626A..86S,Lehmann2024MNRAS.527.4330L, Bloot2025arXiv250214701B}. All but one of the stars in our sample are a few Gyr in age with spin periods of 10--100 days, for which we would expect $B_*$ of order 100s of G \citep{Lehmann2024MNRAS.527.4330L} and a surface X-ray flux of $\lesssim10^{6}$~erg~cm$^{-2}$~s$^{-1}$ or $\lesssim100$ times the solar value \citep{Wright2018MNRAS.479.2351W}. From Figures 1 and 4 of \cite{Bloot2025arXiv250214701B}, we note that our stars would have $\dot M_{\rm wind}$ within an order of magnitude of $\sim10^{-14}$~M$_\odot$~yr$^{-1}$. To account for the uncertainty in stellar magnetospheric properties, we calculate the Alfv\'en radius of each exoplanet by sampling from a log-normal $B_*$ prior with a mean of $200$~G and a width of $\pm0.5$~dex, and likewise a log-normal $\dot M_{\rm wind}$ prior with a mean $\log(\dot M_{\rm wind} [\textrm{M}_\odot~\textrm{yr}^{-1}])$ of $-14$ and a width of $\pm0.5$~dex. We build a posterior of Alfv\'en radii values from these samples via Equation \ref{eqn:AlfvenR} (an example of which is presented in Appendix \ref{sec:alfvenposterior}) and present the mean estimated Alfv\'en radii with uncertainties in Figure \ref{fig:alfvenradius}. Through this estimation, we find $\sim30$ terrestrial exoplanets which orbit within the Alfv\'en radius of their host star at $a\leq R_A$, and therefore may be promising candidates for building circumstellar plasma tori.

Examining the situation in reverse, the sensitivity of the Alfv\'en radius to the stellar magnetospheric properties also suggests that plasma tori can act as exquisite probes of the stellar magnetosphere. By Equation \ref{eqn:etastar}, $\eta_*\propto B_*^2$ suggests that small variations in $B_*$ propagate into much larger and readily-detectable variations in $R_A$. As will be shown in Section \ref{sec:steadystate}, the torus mass itself is $\propto R_A^2$ and therefore changes of $B_*$ by factors of $\sim2$ can yield order-of-magnitude changes in the quasi-steady mass. Thus, measurements of torus quasi-steady masses will place tight constraints on $B_*$. $\eta_*$ is less sensitive to the stellar wind mass loss rate, but we expect that order-of-magnitude constraints on $\dot M_{\rm wind}$ will be possible through quasi-steady torus mass measurements.

\subsection{Confinement Against Radiation Pressure and Stellar Activity} \label{sec:radiation}
The primary difference between the Jupiter-Io system and the proposed analogous star-planet system is the luminosity of the central object. While Jupiter emits a non-trivial radiative flux due to Kelvin-Helmholtz contraction, the luminosity produced by a star through fusion is several orders of magnitude higher. Stellar radiation could exert significant radiation pressure on a circumstellar plasma torus, which in turn might disperse the escaping outgassed material before it can become entrained in a torus. It is therefore crucial to investigate the relative importance of stellar radiation against the confinement of the magnetosphere.

The magnetic pressure experienced by the plasma, under the same dipolar approximation used in Equation \ref{eqn:bfield}, is
\begin{equation} \label{eqn:magp}
    P_{\rm mag}(r)=\frac{B_*^2}{2\mu_0}\bigg(\frac{r}{R_*}\bigg)^{-6}
\end{equation}
The radiation pressure exerted by a star which acts against the magnetic pressure is given by
\begin{equation} \label{eqn:radp}
    P_{\rm rad}(r)=\kappa \Sigma\frac{\sigma_{\rm SB} T_{\rm eff}^4}{c}\bigg(\frac{r}{R_*}\bigg)^{-2}\,.
\end{equation}
In Equation \ref{eqn:radp}, $\kappa$ is the opacity, $\Sigma$ is the surface mass density, and $\sigma_{\rm SB}$ is the Stefan-Boltzmann constant. $\kappa$ is wavelength-dependent and sensitive to the composition, density, and temperature of the torus \citep[e.g.,][]{Semenov2003AA...410..611S}. Under conditions like those in the Io plasma torus \citep[$T\gtrsim1000$~K, $\rho\simeq10^{-19}$~g~cm$^{-3}$; from][]{Delamere2003JGRA..108.1276D} we can estimate a Planck mean opacity of $\kappa\simeq10^5$~cm$^2$~g$^{-1}$ \citep{Semenov2003AA...410..611S}, appropriate for use in optically thin, very hot plasma. The column density of the Io plasma torus \citep[$\lesssim10^{14}$~cm$^{-2}$; from][]{Steffl2004Icar..172...78S} combined with the mass of e.g. a sulfur atom (32 amu) implies a surface mass density of order $\Sigma\lesssim10^{-9}$~g~cm$^{-2}$. Together, these imply an upper bound $\kappa\Sigma\lesssim10^{-3}$. The ratio of Equations \ref{eqn:magp} and \ref{eqn:radp} yields a ``radiation magnetic confinement parameter'' which we denote with the symbol $\zeta$:
\begin{equation} \label{eqn:pratio}
    \zeta(r)=\zeta_*\bigg(\frac{r}{R_*}\bigg)^{-4}\,.
\end{equation}
In Equation \ref{eqn:pratio}, $\zeta_*$ is the radiation magnetic confinement parameter at the stellar surface:
\begin{equation} \label{eqn:zetastar}
\zeta_*=\underbrace{\bigg(\frac{B_*^2}{2\mu_0}\bigg)}_\textrm{Magnetic Field}\,\underbrace{\bigg(\kappa\Sigma\frac{\sigma_{\rm SB} T_{\rm eff}^4}{c}\bigg)^{-1}}_\textrm{Radiation}\,.
\end{equation}

\begin{figure}[h!]
    \centering
    \includegraphics[width=1.00\linewidth]{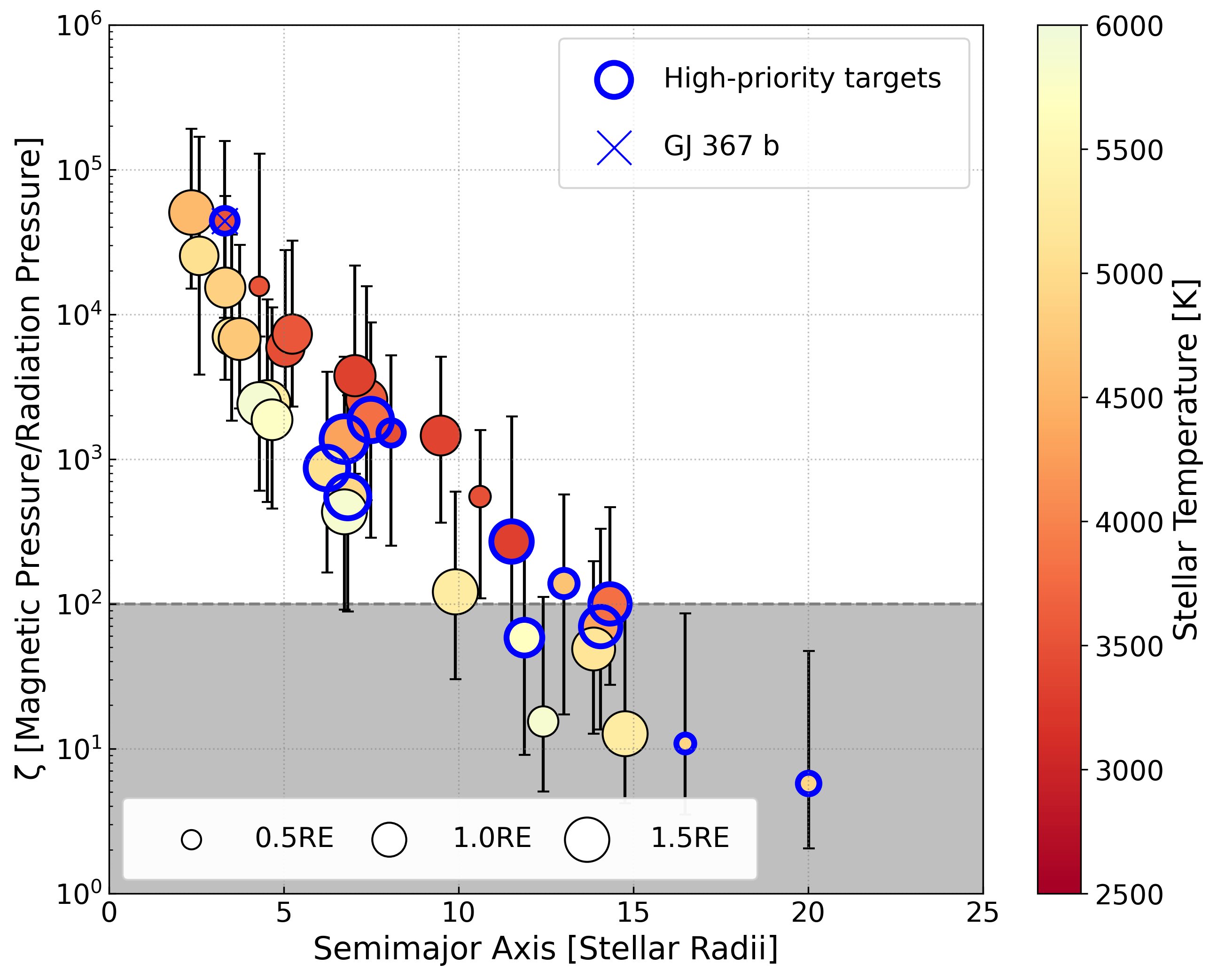}
    \caption{Radiation magnetic confinement parameter $\zeta$ for the 33 exoplanets which orbit at $a<R_A$ identified in Figure \ref{fig:alfvenradius}. For all but eight of these targets, the magnetic pressure is at least two orders of magnitude greater than the radiation pressure, such that radiation pressure is unlikely to be capable of dispersing magnetically-confined plasma. Targets in the grey region have $\zeta\leq100$ such that their tori may be dispersed by radiation pressure.}
    \label{fig:zeta}
\end{figure}

In Figure \ref{fig:zeta}, we show $\zeta(a)$ for all of the exoplanets in Figure \ref{fig:alfvenradius} which orbit within the Alfv\'en radius, using $\kappa\Sigma\lesssim10^{-3}$. $\zeta(a)$ is estimated using the same $B_*$ prior as was used in estimating the Alfv\'en radius. We find that all but eight of these exoplanets have $\zeta(a)\geq100$ such that magnetic pressure far outweighs the effects of radiation pressure. For targets with $\zeta(a)\geq100$, we neglect the possibility of torus erosion by quiescent stellar radiation. For $\zeta(a)<100$ targets, radiation pressure may disperse plasma, reducing the torus quasi-steady mass or altogether preventing torus formation.

Transiting rocky exoplanets are frequently found around low-mass G, K, and M stars because small planets are easier to detect around smaller stars. These low-mass stars are prone to flaring. As a well-known example, the fast-rotating M dwarf star TRAPPIST-1 \citep[$P_*=3.295$~days; from][]{Vida2017ApJ...841..124V} experiences tens of flares per day \citep{Vida2017ApJ...841..124V}, including multiple flares per day at a level of $\gtrsim100\%$ of the quiescent luminosity \citep{Howard2023ApJ...959...64H} and multiple ``superflares'' per year at $\gtrsim10^{33}$~erg~s$^{-1}$ \citep{Glazier2020ApJ...900...27G}. Coronal mass ejections (CMEs) associated with these flares travel faster than the typical stellar windspeed and can enhance the stellar wind mass loss rate by an order or two of magnitude \citep{Vida2016AA...590A..11V}. Since $\eta_*$ scales inversely with both $v_\infty$ and $\dot M_{\rm wind}$, flares and associated CMEs could cause localized reductions in $\zeta(r)$ and $\eta(r)$ by orders of magnitude. This will unbind the plasma in flare-stricken regions of the torus, opening gaps or fully destroying the torus if flaring is frequent enough. It is therefore crucial to determine whether we expect excessive flare activity for any of our targets.

Fast-rotating stars with rotation periods $\lesssim3$~days experience more frequent high-energy flares than slow-rotating stars \citep{Seligman2022ApJ...929...54S}. In contrast, the frequency of flare events and associated CMEs decreases with intensity of the event for stars with rotation periods of $\gtrsim3$~days, regardless of spectral type \citep{Feinstein2022ApJ...925L...9F}. All but one of our targets have spin periods $\geq10$ days. Therefore, the extreme flaring activity associated with fast-rotating M dwarfs like TRAPPIST-1 is unlikely to be found in most of our target systems. The exception is HD~63433~A, which has a rotation period of 6.4~days and a young age of $\lesssim500$~Myr \citep{Capistrant2024AJ....167...54C}, and thus may exhibit much more frequent flaring and CMEs. For the HD~63433 system, it is possible that excessive stellar activity may fully prevent torus formation. For all other systems, we expect flares with $\sim100\%$ of the quiescent stellar luminosity to occur on average every 10--100 days \citep{Feinstein2022ApJ...925L...9F}, of which only some will be traveling in the correct direction to strike the torus. We will show in the next section that circumstellar plasma tori replenishment timescales are often of order weeks to months, so that replenishment acts at rates comparable to the rate of stellar activity-driven erosion. We therefore expect that flare activity will not fully prevent torus formation. Nonetheless, some flare-induced damage is expected in circumstellar plasma tori, for example in the form of gaps torn open by flares. Acquiring multiple observations of the system separated by several months will thus be key to determining whether a torus non-detection was a true negative or was a false negative resulting from observing a flare-opened gap.

We note another opportunity to characterize stellar magnetospheres arises here. Because more frequent flaring drives more significant torus erosion, tori quasi-steady masses can be expected to decline during periods of enhanced stellar activity. The Sun has an activity cycle of 11 years, during which stellar activity gradually ramps up to a maximum before a large-scale reconfiguration of the solar magnetic field resets the activity back to lower levels \citep[e.g.,][]{Wu2018AA...620A.120W, Owens2021SoPh..296...82O}. Long-term monitoring of circumstellar plasma tori may reveal cyclical changes in quasi-steady mass. Such changes would be directly correlated with variations in stellar magnetic activity and would allow the periodicity of stellar magnetic cycles to be determined.

\subsection{Timescale of Torus Dispersal} \label{sec:timescale}
A plasma torus has equal rates of mass injection and removal in equilibrium. In reality the mass-loss is not instantaneous, which permits a quasi-steady equilibrium torus mass $M_{\rm torus}$ to accumulate. The removal timescale of newly-injected mass is the magnetospheric convection timescale, $\tau$. The torus mass then depends on the product of the mass injection rate and $\tau$.

\begin{figure}[h!]
    \centering
    \includegraphics[width=1.00\linewidth]{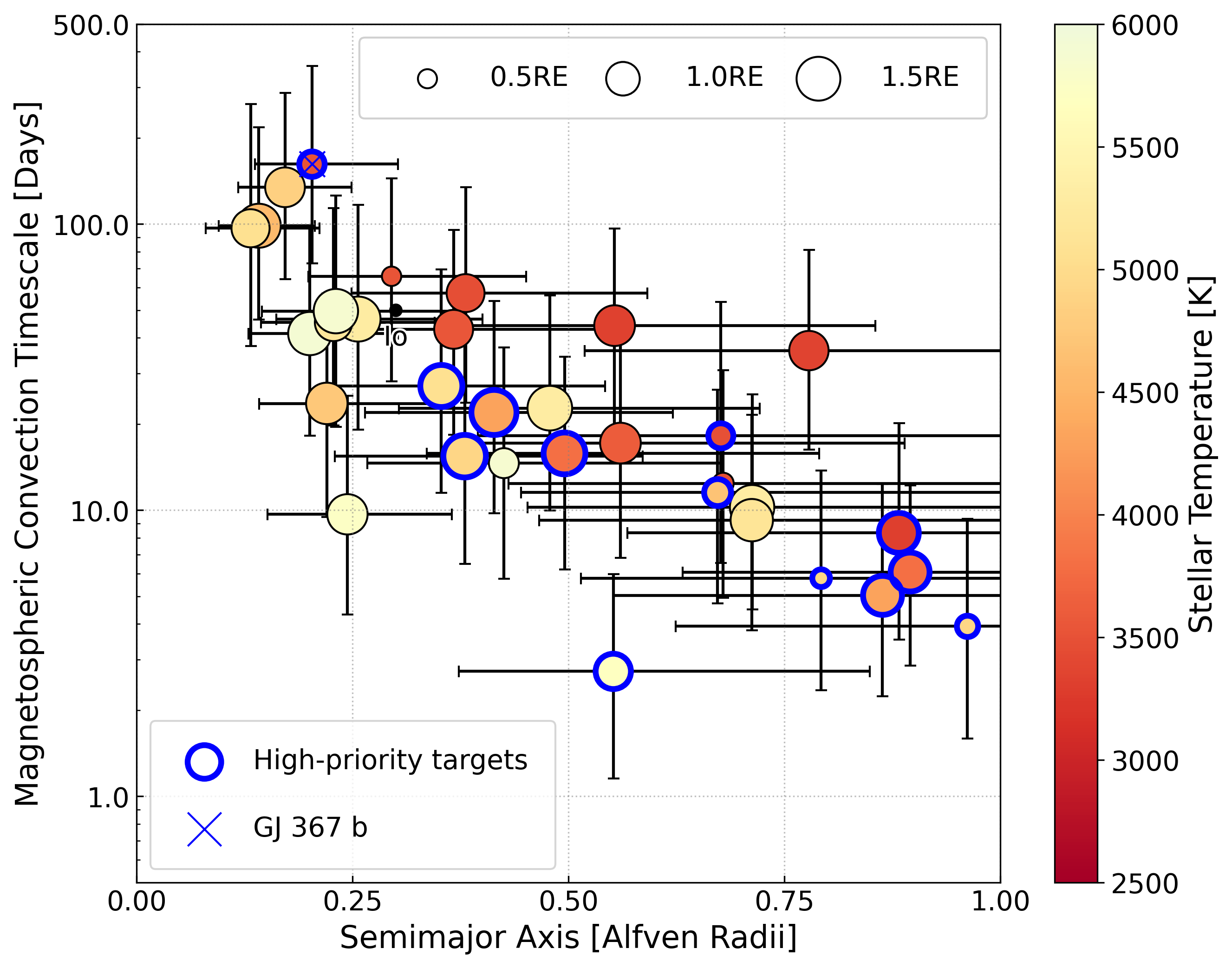}
    \caption{Magnetospheric convection timescale $\tau$ over which mass injected into a plasma torus by an exoplanet at distance $a/R_A$ would be removed. Less time is required to remove mass when the injection occurs closer to the Alfv\'en radius. We calculate $\tau$ with Equation \ref{eqn:timescale} using the same assumptions and exoplanets as in Figure \ref{fig:alfvenradius}, restricting the plot to show only exoplanets which orbit at $a<R_A$. The value of $\tau$ for Io is sourced from \cite{BagenalDols2020}.}
    \label{fig:timescale_lols}
\end{figure}

Magnetospheric convection is not fully understood even in the well-studied Jupiter-Io system. \cite{Hill1981} predicted $\tau\sim14$~hours, which underestimates the measured Jupiter-Io system timescale by more than an order of magnitude \citep[20--80 days; see e.g.][]{BagenalDols2020}. This underestimation is likely due to processes that delay magnetospheric convection including ring current impoundment and velocity shear \citep{Thomas2004jpsm.book..561T}. As a first estimate we use Equation 24 of \cite{Hill1981} as a lower bound on the timescale; we will discuss future work on this topic in Section \ref{sec:caveats}. We adapt Equation 24 of \cite{Hill1981} to our star-planet system, assuming that material enters the torus at the planet orbit semimajor axis $a$ and escapes from the torus at the stellar Alfv\'en radius $R_A$:
\begin{equation} \label{eqn:timescale}
    \tau\gtrsim\,\sqrt{\frac{2}{3}}\,\,\bigg(\frac{R_A}{a}\,\bigg)^2\,\bigg(\,\frac{1}{\Omega}\,\bigg)\,.
\end{equation}
By estimating the stellar Alfv\'en radius through Equation \ref{eqn:AlfvenR}, we use Equation \ref{eqn:timescale} to obtain a lower bound on the magnetospheric convection timescale and equivalently place a lower bound on how long material remains in the torus before being dispersed by the stellar wind. In Figure \ref{fig:timescale_lols}, we show how $\tau$ scales with the ratio of the orbit distance to the Alfv\'en radius for the exoplanets in Figure \ref{fig:alfvenradius} that orbit interior to their host star's Alfv\'en radius.

\subsection{The Geometry of the Torus} \label{sec:span}
In the Io plasma torus, mass injection occurs at Io's orbit distance, and mass escape occurs at the Jovian Alfv\'en radius. However, the material in the Io plasma torus is not distributed evenly between these two radial distances. Instead, $\sim90\%$ of the torus mass is concentrated within the ``warm torus'', the component of the torus which spans $6$~R$_{\rm J}$ to 9~R$_{\rm J}$ \citep{BagenalDISTRO1980GeoRL...7...41B,BagenalDols2020}. Note that the Jovian Alfv\'en radius is located at $\sim20$~R$_{\rm J}$ \citep{Hill1981}, considerably further than the outer edge of the warm torus. We therefore assume that in a circumstellar torus, mass injection occurs at the exoplanet orbit semimajor axis $a$, mass escape occurs at the stellar Alfv\'en radius $R_A$, and plasma is not evenly distributed between these radial distances.

We approximate the thickness of the torus as follows. Torus plasma corotating at a rate $\Omega$ diffuses latitudinally from the torus centrifugal equator following an exponentially-decaying diffusion law with scale height defined in \cite{Bagenal1994JGR....9911043B} as
\begin{equation} \label{eqn:H}
    H=\sqrt{\frac{2k_BT}{3m\Omega^2}}\,.
\end{equation}
In Equation \ref{eqn:H}, $k_B$ is the Boltzmann constant and $T$ and $m$ are the temperature and mass of the plasma ions. The latter is estimated using the equilibrium temperature $T_{\rm eq}$ of a zero-albedo absorber at the planetary orbit distance $a$. Note that the temperature estimated this way is typically 1000--2000~K, comparable in order of magnitude to the temperature of Io’s ionosphere as a result of plasma heating \citep{Strobel1994}. We assume that the plasma torus is isotropic such that plasma density decays with radial distance $r$ following the same exponentially-decaying diffusion law with the same scale distance $H$. For an exponentially-decaying density, $\sim95\%$ of the mass is contained within three scale radii; as this is comparable to the $\sim90\%$ of the mass contained within the warm torus, we assume that the thickness of the circumstellar plasma tori is three scale radii.

\begin{figure*}\centering
    \includegraphics[width=1.00\linewidth]{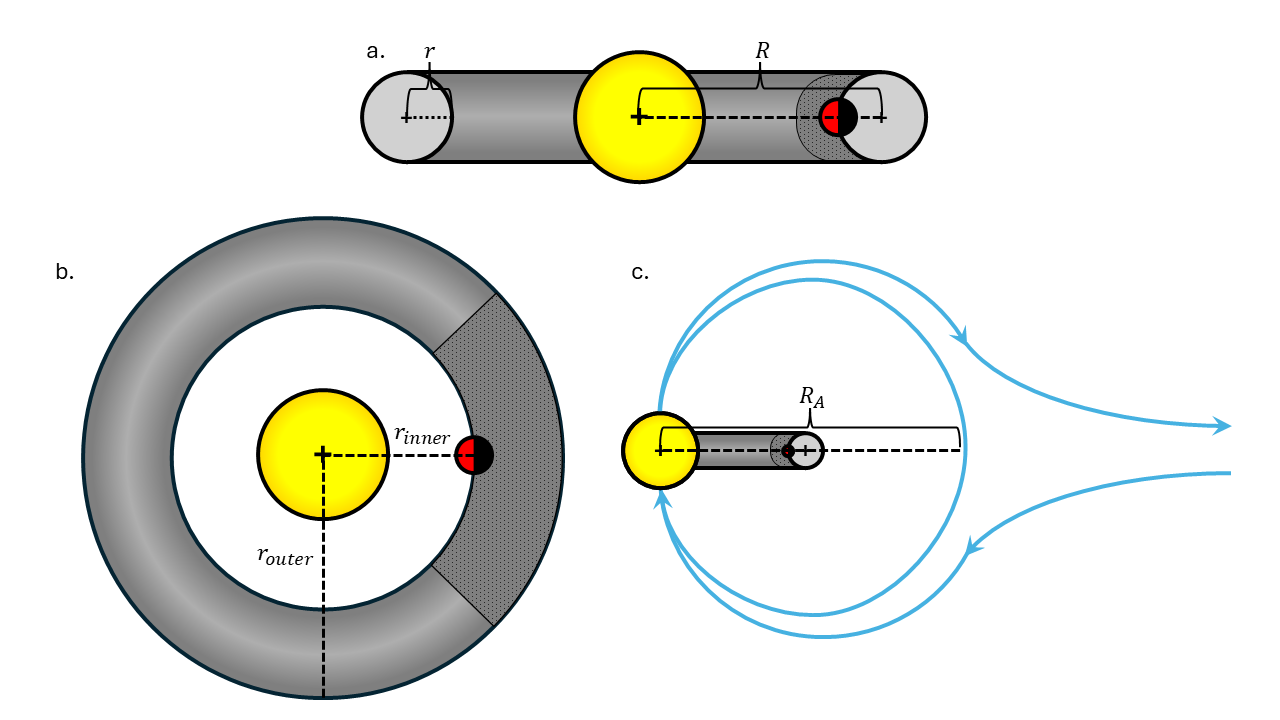}
    \caption{Schematic illustration of the plasma torus (grey) built by a volcanically-active exoplanet (red shaded circle) in orbit around a star (yellow circle), with a neutral cloud extending a few planetary radii ahead and behind the source exoplanet (dotted grey). a. The torus major radius $R=a+(3H/2)$ is the distance from the star's center to the center of the tube. The minor radius $r=3H/2$ is half the diameter of the tube. b. The inner radius $r_{\rm inner}=a$, the planet orbit distance. The outer radius $r_{\rm outer}=a+3H$ marks the edge of the warm torus and contains 90\% of the torus mass. c. The Alfv\'en radius $R_A$ typically lies much further out than $r_{\rm outer}$. The stellar magnetic field loops (in blue) switch from a closed to a radially outwards and open geometry \citep{ud2002ApJ...576..413U} at the Alfv\'en radius.}
    \label{fig:torus}
\end{figure*}

Therefore, the inner edge of the plasma torus, $r_{\rm inner}$, begins at the mass injection distance such that $r_{\rm inner}=a$. The outer edge of the torus, $r_{\rm outer}$, is located three scale heights further such that $r_{\rm outer}=a+3H$. In Figure \ref{fig:torus}, we show a schematic illustration of our fiducial plasma torus model with relevant dimensions indicated. This fiducial torus has a total volume, $V_{\rm torus}$, given by
\begin{equation} \label{eqn:volume}
    V_{\rm torus}\simeq\frac{9}{2}\pi^2 H^2 a\bigg(1+\frac{3H}{2a}\bigg)\,.
\end{equation}
While not accounting for smaller components of the torus such as the cold disk or ribbon seen in the Jupiter-Io system, the warm torus encompasses $\sim90\%$ of Io's torus mass \citep{BagenalDISTRO1980GeoRL...7...41B}. Therefore, our approximate geometry in Equation \ref{eqn:volume} should also encompass $\sim90\%$ of the torus mass, sufficient for an order of magnitude estimate. Further discussion of how torus geometry assumptions affect these results is deferred to Section \ref{sec:caveats}.

\subsection{The Steady State Torus Mass} \label{sec:steadystate}
In this subsection we calculate the steady state torus mass using the assumptions outlined in the previous subsections. The steady state mass is approximately equal to the mass injection rate multiplied by the magnetospheric convection timescale, $M_{\rm torus}\simeq\tau \dot M_{\rm torus}$. The mass injection rate will be equal to the fraction of outgassed volatiles that escape from the planet, $\dot M_{\rm torus}=f_{\rm esc}\dot M_{\rm volc}$. By combining Equation \ref{eqn:AlfvenR} with Equation \ref{eqn:timescale}, we calculate a lower limit on the magnetospheric convection rate and steady state mass for the torus:
\begin{equation} \label{eqn:torusmass}
    M_{\rm torus} \gtrsim \sqrt{\frac{2}{3}}f_{\rm esc} \bigg(\frac{\dot M_{\rm volc}}{\Omega}\bigg)\bigg(\frac{R_A}{a}\bigg)^2\,.
\end{equation}
In the next section, we will use Equation \ref{eqn:torusmass} to show how the volcanic outgassing rate $\dot M_{\rm volc}$ can be estimated from observable quantities, and what observational strategies can be employed to characterize and constrain the detected torus.

\section{Torus Mass Estimates from Observables and Observational Strategy} \label{sec:analysis}
We consider a transiting exoplanetary system where the star, planet, and torus are viewed edge-on. In this configuration the torus perpetually occults the host star and imprints absorption features on the stellar spectrum at wavelengths dependent on its composition.

\subsection{Inferring Torus Mass from Absorption Features} \label{sec:absorption}
We consider a hypothetical absorption signature with strength $f_{\rm abs}$. This absorption strength depends on the number density $n$ (assumed to be constant within the torus) and cross-section $\sigma$ of absorbers along the line of sight. Recall that we assume a torus thickness of $\simeq3H$. Therefore, the column density of absorbers along the line of sight is approximately $3Hn$ and the absorption strength will be
\begin{equation} \label{eqn:abs}
    f_{\rm abs} \simeq 3Hn\,\sigma \,.
\end{equation}
The torus plasma is sufficiently rarefied to act as an ideal MHD fluid. Therefore, the plasma is not collisionally dominated, reflected in the fact that Equation \ref{eqn:H} depends on the mass of the torus particles. Each absorbing species (denoted with subscript $i$) will distribute across its own thickness $H_{\rm i}$ and occupy a species-specific volume $V_{\rm torus,\,i}$. For the remainder of this section, we treat each species as occupying its own unique torus. Each species-specific torus will have a total mass based on the species' particle mass and number density, $M_{\rm torus,\,i}=n_{\rm i}m_{\rm i}V_{\rm torus,\,i}$. Furthermore, each species imprints its own absorption feature with strength $f_{\rm abs,\,i}$.

We relate the species-specific torus mass to its absorption feature as follows. Substituting in the volume from Equation \ref{eqn:volume} and the number density obtained from rearranging Equation \ref{eqn:abs} yields
\begin{equation} \label{eqn:torusmassspecies}
    M_{\rm torus,\,i} \simeq \frac{9}{4}\pi^2f_{\rm abs,\,i} \bigg(\frac{H_{\rm i}^2}{\sigma_{\rm i}}\bigg) \bigg(1+\frac{2a}{3H_{\rm i}}\bigg)m_{\rm i}\,.
\end{equation}
Recall that the torus mass is constrained with Equation \ref{eqn:torusmass}. Combining that result with Equation \ref{eqn:torusmassspecies} and rearranging, we obtain the following inequality for the mass injection rate of species $i$, $f_{\rm esc,\,i}\dot M_{\rm volc,\,i}$:
\begin{equation} \label{eqn:mvolc1}
\begin{split}
    f_{\rm esc,\,i}\dot M_{\rm volc,\,i} \lesssim \bigg(\frac{3}{2}\bigg)^{5/2}\pi^2 f_{\rm abs,\,i}\bigg(\frac{H_{\rm i}^2}{\sigma_{\rm i}}\bigg)\bigg(\frac{a}{R_A}\bigg)^2\\ \times\bigg(1 + \frac{2a}{3H_{\rm i}}\bigg)\, m_{\rm i}\Omega\,.
    \end{split}
\end{equation}

\subsection{Observing Strategies for Torus Characterization} \label{sec:strategy}
We have shown how the volcanic outgassing rate can be linked to detectable absorption features. We next outline observing strategies by which torus absorption features may be observed so that Equation \ref{eqn:mvolc1} or analogous equations may be applied.

\subsubsection{Neutral Absorption Signatures}
Molecules and their photodissociated byproducts escape from the atmosphere of the source planet in a largely neutral state, and form an extended cloud ahead and behind the planet in its orbit. If the cloud particles remain neutral, the stellar magnetic field will have no confinement over them, and the stellar wind will eventually blow them away. The timescale over which these particles either ionize or escape to the stellar wind, combined with the volcanic outgassing rate, sets the total mass of the neutral cloud. The volume of the neutral cloud is set by the velocity at which the material diffuses ahead of and behind the planet in its orbit, as well as by the timescale of ionization and stellar wind escape. Therefore, Equation \ref{eqn:mvolc1} cannot be used as-is to relate neutral cloud column density back to volcanic outgassing, since (i) the volume spanned by the neutral cloud is considerably smaller than the full torus volume, and (ii) the timescale $\tau$ used to derive Equation \ref{eqn:mvolc1} assumes that the dispersal of the observed mass is driven by magnetospheric convection, a process which does not operate on neutral material. An analogous equation with an updated volume and timescale would need to be derived and employed for this purpose.

Because the neutral cloud has a finite extent, its observation is phase-dependent, only being visible during and around the transit of the source planet. Studies of the neutral cloud could thus be carried out the same way as studies of escaping atmospheres, such as the Lyman-$\alpha$ transit ingress and egress used to study the neutral hydrogen cloud around GJ~3470~b \citep{Bourrier2018...620A.147B}. Likewise, neutral cloud column density as a function of distance from the exoplanet would be determined by observing the transit and its extended ingress and egress. The column density as a function of distance from the planet could then be integrated over the volume of the neutral cloud to infer the total mass of neutral atoms and molecules present in the cloud. The short duration of transit for exoplanets orbiting within the magnetosphere ($P_{\rm Orb}\lesssim10$~days typically produces transit durations $\lesssim2$ hours) makes it possible to observe frequent transits of the neutral cloud with only a few hours of time on space-based observatories, making this method very accessible to modern instrumentation.

It is important to note that neutral cloud detection, while possibly an avenue for confirming extrasolar volcanism, cannot be used to confirm the existence of a circumstellar plasma torus. The formation of a distended cloud of neutral escaping atoms can be achieved fully independent of magnetic confinement. Therefore, signatures from the full torus must be pursued to prove its existence.

While the torus composition is expected to be predominantly ionized material due to the reliance of torus formation on magnetic confinement, there exists a means by which neutrals could disperse throughout the torus. Neutral sodium atoms have been found throughout the entire volume of the Io plasma torus \citep{Brown1974ApJ...187L.125B,Trafton1974ApJ...190L..85T,Bergstralh1975ApJ...195L.131B} despite not being subject to magnetic confinement. The neutral sodium atoms disperse by acting as passengers on ionized molecules, e.g. NaS$^+$, which are swept throughout the torus volume by magnetospheric corotation before dissociating to yield a neutral sodium atom \citep{Johnson1994Icar..111...65J}. Therefore, the possibility of observing toroidal neutrals exists, although it is much harder to quantify the expected neutral density \textit{a priori}. This is in part due to uncertainties in the neutral supply rate, resulting from the many mechanisms by which the carrier ionized molecules can form \citep{Johnson1993GeoRL..20.1735J}. It is furthermore unclear how long such neutrals can remain in the torus before (i) being swept away by stellar winds, (ii) recombining with an ion to form a magnetically-confined molecular ion, or (iii) being ionized themselves by either stellar XUV flux or by collisions with corotating plasma. We opt to relate neutral absorption features to neutral supply rates through Equation \ref{eqn:mvolc1}, but we recommend that future studies explore these neutral seeding and removal processes in more detail to determine the limitations of applying Equation \ref{eqn:mvolc1} to neutral atom absorption signatures.

\subsubsection{Ionic Absorption Signatures}
If the neutral cloud particles become ionized, which is apt to happen due to collisions with energetic torus ions (as discussed in Section \ref{sec:supply}), the stellar magnetosphere may confine them and drag them throughout the orbit path. If there are no holes due to flare- and CME-driven erosion, the torus will create a set of constant ionic absorption signatures that are observable at all exoplanetary orbit and stellar rotation phases. Assuming the torus has azimuthally uniform density to an order of magnitude, periodic ``snapshot'' observations of the star while the planet is not in transit will allow the observer to measure the torus-contaminated stellar spectrum. By phase folding these observations with the stellar rotation period, we may constrain $n$ and estimate $\dot M_{\rm volc}$ through Equation \ref{eqn:mvolc1} for any ionic species present in the torus.

In the next section, we apply Equation \ref{eqn:mvolc1} to find known exoplanets which are most likely to host detectable circumstellar plasma tori, and discuss how these observing strategies will apply to these targets.

\section{High-priority Targets for Circumstellar Plasma Torus Detection} \label{sec:results}
Exoplanets most likely to build substantial plasma tori around their host stars will have (i) small semimajor axes $a$ relative to the stellar Alfv\'en radius $R_A$ in order to maximize $\tau$, (ii) a significant volcanic outgassing rate $\dot M_{\rm volc}$, and (iii) an appreciably nonzero escape fraction $f_{\rm esc}$. We narrow our focus here to terrestrial exoplanets on eccentric orbits. These exoplanets are susceptible to tidal heating which can generate vigorous and sustained outgassing over Gyr timescales, and therefore we consider them high-priority targets for future study. The tidal heating power experienced by eccentric exoplanets was derived in \cite{Beuthe2013Icar..223..308B} based on the work of \cite{Peale1978Icar...36..245P} and \cite{Peale1979Sci...203..892P}. We write it here in the scaled form used in \cite{Seligman2024ApJ...961...22S}:
\begin{equation} \label{eqn:tides}
\begin{split}
  \dot E_{heat} = (3.4\times10^{25} \textrm{ erg s}^{-1})\,\bigg(\frac{P_{\rm Orb}}{1\textrm{ d}}\bigg)^{-5} \\ \times\,\bigg(\frac{R}{R_\oplus}\bigg)^5\,\bigg(\frac{e}{10^{-2}}\bigg)^2 \,\bigg(\frac{\Im(\Tilde{k_2})}{10^{-2}}\bigg)\,.
\end{split}
\end{equation}
In Equation \ref{eqn:tides}, $\Im(\Tilde{k_2})$ is the imaginary part of the Love number, equivalent in some rheologies to the inverse tidal quality factor $1/Q$ and characteristic of how the planetary bulk responds to deformation.

\begin{figure}[h!]
  \centering
  \includegraphics[width=1.00\linewidth]{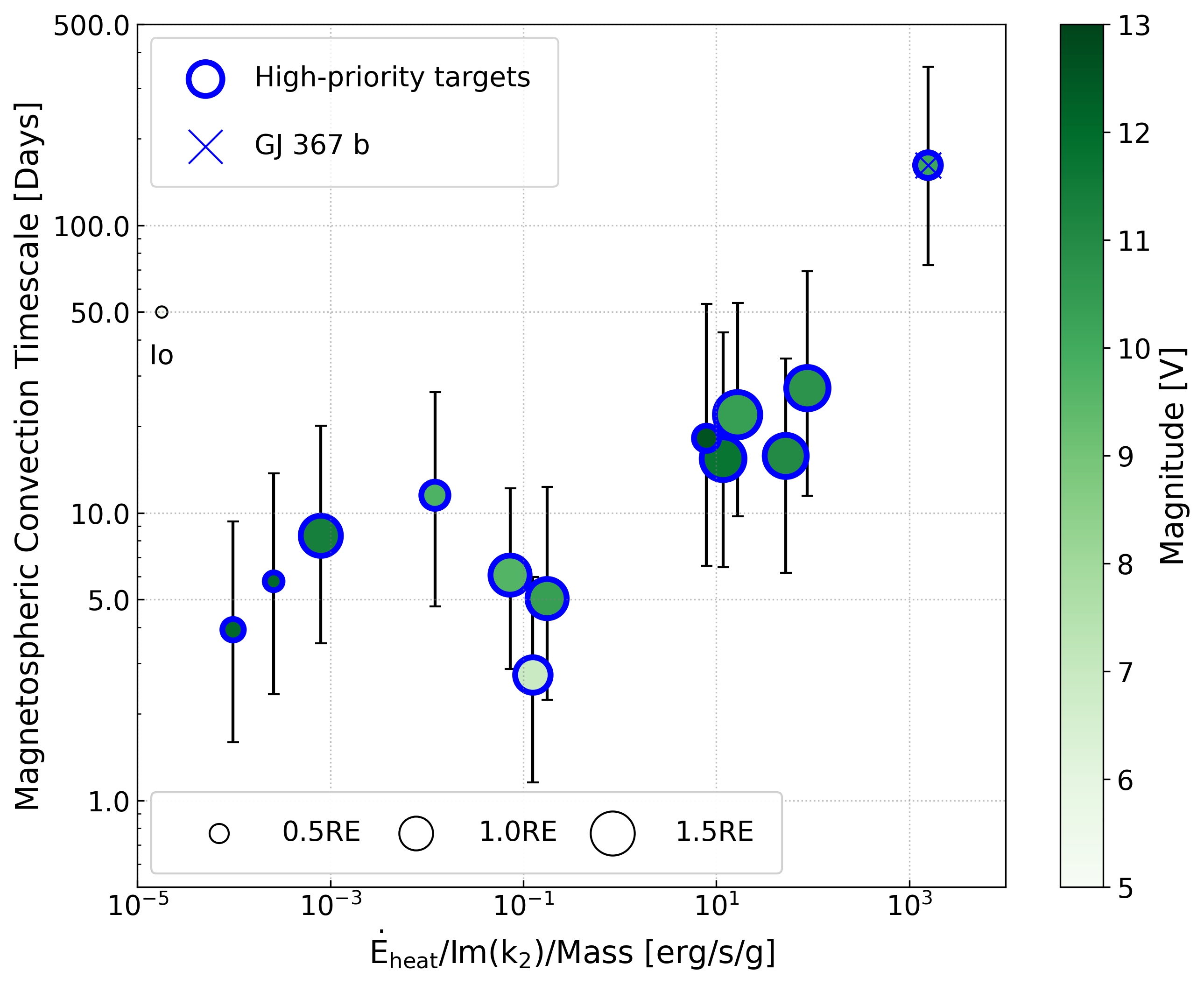}
  \caption{Magnetospheric convection timescale shows a positive correlation with the tidal heating rate. This suggests that exoplanets which are well-positioned for plasma torus construction will generally also receive the necessary tidal power to drive volcanism. This plot includes the 13 terrestrial exoplanets plotted in Figures \ref{fig:alfvenradius} and \ref{fig:timescale_lols} which had nonzero orbit eccentricities and orbited within their host stars' Alfv\'en radii, marked on all figures as ``high-priority targets''. GJ~367~b is considered the highest priority target due to its high tidal heating power and long magnetospheric convection timescale. The values of $\tau$ presented here are the same as in Figure \ref{fig:timescale_lols}. Color indicates the V magnitude of the system which is indicative of favorability for UV-optical spectroscopy.}
  \label{fig:tides}
\end{figure}

Figure \ref{fig:tides} shows a positive correlation between an exoplanet's magnetospheric convection timescale $\tau$ and its tidal heating rate $\dot E_{\rm heat}$. This relationship is expected as both $\tau$ and $\dot E_{\rm heat}$ increase with decreasing orbit semimajor axes $a$ and periods $P_{\rm Orb}$. The exoplanets included in Figure \ref{fig:tides} are the subset of exoplanets in Figures \ref{fig:alfvenradius} and \ref{fig:timescale_lols} which were marked as ``high-priority targets'', receiving this designation because they had $e>0$ and $a<R_A$. In Table \ref{tab:results}, we summarize our high-priority targets for circumstellar plasma tori detection, organizing targets by those with the largest values of $\tau$. We propagate the uncertainties on $R_A$ and $\tau$ derived in Figures \ref{fig:alfvenradius} and \ref{fig:timescale_lols} into Figure \ref{fig:tides} and Table \ref{tab:results}.

\setlength{\tabcolsep}{1.5pt}
\begin{table*}[t]
\begin{rotatetable*}
\begin{center}
\caption{Exoplanets with properties that are amenable to building plasma tori}\label{tab:results}
\begin{tabular}{ ccccccccccccccccc} 
\hline
\hline
Name & a & R$_\textrm{A}$ & H & $\zeta$ & P$_\textrm{Orb}$ & $e$ & R$_{\rm P}$ & T$_\textrm{eq}$ & M$_\star$ & R$_\star$ & P$_\star$ & V & D & $\dot{\rm E}_\textrm{heat}/\Im(\tilde{k}_2)$ & $\tau$ & $e$ Ref. \\
&[au]&[au]&[10$^5$ km]&&[days]&&[R$_\oplus$]&[K]&[M$_\odot$]&[R$_\odot$]&[days]&[Mag]&[pc]&[erg s$^{-1}$]&[days] & \\
\hline
GJ 367 b & 0.007 & 0.035$^{+0.017}_{-0.012}$ & 3.43 & 44252 & 0.322 & 0.060$^{+0.070}_{-0.040}$ & 0.699 & 1368 & 0.46 & 0.46 & 51.30 & 10.153 & 9 & 5.91E+30 & 162$^{+194}_{-89}$ & RV; [1]\\ 
TOI-238 b & 0.021 & 0.060$^{+0.036}_{-0.021}$ & 1.78 & 867 & 1.273 & $<$0.180 & 1.402 & 1432 & 0.79 & 0.73 & 26.00 & 10.748 & 81 & 1.78E+30 & 27$^{+42}_{-16}$ & RV; [2]\\ 
GJ 9827 b & 0.019 & 0.045$^{+0.026}_{-0.015}$ & 1.79 & 1374 & 1.209 & $<$0.063 & 1.577 & 1173 & 0.61 & 0.60 & 28.90 & 10.370 & 30 & 5.10E+29 & 22$^{+32}_{-12}$ & RV; [3]\\ 
LHS 1678 b & 0.012 & 0.018$^{+0.013}_{-0.007}$ & 3.41 & 1516 & 0.860 & 0.033$^{+0.035}_{-0.023}$ & 0.685 & 869 & 0.34 & 0.33 & 64.00 & 12.600 & 20 & 1.19E+28 & 18$^{+35}_{-12}$ & RV; [4]\\ 
L 168-9 b & 0.021 & 0.042$^{+0.020}_{-0.016}$ & 1.69 & 1863 & 1.401 & $<$0.210 & 1.390 & 982 & 0.62 & 0.60 & 29.80 & 11.005 & 25 & 1.44E+30 & 16$^{+19}_{-10}$ & RV; [5]\\ 
K2-36 b & 0.023 & 0.060$^{+0.040}_{-0.021}$ & 1.13 & 549 & 1.423 & $<$0.093 & 1.430 & 1330 & 0.79 & 0.72 & 17.13 & 11.726 & 110 & 3.02E+29 & 15$^{+27}_{-9}$ & RV; [3]\\ 
HD 23472 d & 0.043 & 0.064$^{+0.033}_{-0.023}$ & 2.20 & 138 & 3.977 & 0.070$^{+0.050}_{-0.047}$ & 0.750 & 918 & 0.67 & 0.71 & 40.10 & 9.730 & 39 & 3.98E+25 & 12$^{+15}_{-7}$ & RV; [6]\\ 
GJ 486 b & 0.017 & 0.019$^{+0.011}_{-0.007}$ & 2.37 & 270 & 1.467 & 0.001$^{+0.002}_{-0.000}$ & 1.289 & 691 & 0.31 & 0.32 & 49.90 & 11.390 & 8 & 1.32E+25 & 8$^{+12}_{-5}$ & T; [7], [8]\\ 
HD 260655 b & 0.029 & 0.033$^{+0.014}_{-0.010}$ & 1.81 & 100 & 2.770 & 0.039$^{+0.043}_{-0.023}$ & 1.240 & 710 & 0.44 & 0.44 & 37.50 & 9.630 & 10 & 9.30E+26 & 6$^{+6}_{-3}$ & RV; [9]\\ 
Kepler-102 b & 0.055 & 0.070$^{+0.038}_{-0.025}$ & 1.48 & 11 & 5.287 & $<$0.100 & 0.460 & 855 & 0.80 & 0.72 & 27.95 & 12.072 & 108 & 1.70E+24 & 6$^{+8}_{-3}$ & RV; [3]\\ 
GJ 9827 c & 0.039 & 0.045$^{+0.026}_{-0.015}$ & 1.49 & 70 & 3.648 & $<$0.094 & 1.241 & 812 & 0.61 & 0.60 & 28.90 & 10.370 & 30 & 1.37E+27 & 5$^{+7}_{-3}$ & RV; [3]\\ 
Kepler-102 c & 0.067 & 0.070$^{+0.038}_{-0.025}$ & 1.41 & 6 & 7.071 & $<$0.094 & 0.567 & 776 & 0.80 & 0.72 & 27.95 & 12.072 & 108 & 9.96E+23 & 4$^{+5}_{-2}$ & RV; [3]\\ 
HD 63433 d & 0.050 & 0.091$^{+0.044}_{-0.032}$ & 0.40 & 58 & 4.209 & 0.160$^{+0.360}_{-0.120}$ & 1.073 & 1167 & 0.99 & 0.91 & 6.40 & 6.920 & 22 & 9.37E+26 & 3$^{+3}_{-2}$ & T; [10]\\ 
\hline
\multicolumn{17}{p{1.1\textwidth}}{\small{\textbf{Notes. }The table shows orbit semimajor axis $a$, Alfv\'en radius $R_A$, torus scale radius $H$, radiation magnetic confinement parameter $\zeta$, orbital period $P_{\rm Orb}$, eccentricity $e$, planetary radius $R_{\rm P}$, planetary equilibrium temperature $T_{eq}$, stellar mass $M_*$, stellar radius $R_*$, stellar rotation period $P_*$, magnitude $V$, system distance $D$, tidal heating $\dot{E}_{\rm heat}/\Im(\tilde{k}_2)$, and magnetospheric convection timescale $\tau$. The value of $H$ presented corresponds to a sulfur particle of mass $m=32$~amu; lighter (heavier) particles will have higher (lower) values of $H$. Note that the value listed $\dot{E}_{\rm heat}/\Im(\tilde{k}_2)$ corresponds to the case of $\Im(\tilde{k}_2)=1$. For reference the tidal heating of Io is $\simeq1.6\times10^{21}/\Im(\tilde{k}_2)$ ~erg~s$^{-1}$. References for $e$ and identification method (RV = radial velocity or radial velocity + transit fits; T = transit fits)  provided in the last column are: [1] \cite{Goffo2023ApJ...955L...3G}, [2] \cite{Suarez2024AA...685A..56S}, [3] \cite{Bonomo2023AA...677A..33B}, [4] \cite{Silverstein2024AJ....167..255S}, [5] \cite{ADef2020AA...636A..58A}, [6] \cite{Barros2022AA...665A.154B}, [7] \cite{WeinerMansfield2024ApJ...975L..22W}, [8] \cite{Caballero2022AA...665A.120C}, [9] \cite{Luque2022AA...664A.199L}, and [10] \cite{Capistrant2024AJ....167...54C}.}} \end{tabular}
\end{center}
\end{rotatetable*}
\end{table*}

We next calculate the level of volcanic activity required to produce plasma torus absorption detectable with HST. Previous studies collecting exoplanet transmission spectra with HST using STIS \citep{Bourrier2013...551A..63B, Bourrier2018...620A.147B} and COS \citep{DosSantos2023AJ....166...89D} were able to detect atomic Lyman-$\alpha$ absorption of order 10\%. Similarly, HST STIS E230M Echelle grating surveys searching for escaping Mg~II measured transit changes in flux at the Mg~II doublet ($\sim280$~nm) of 5--10\% to a confidence of 5$\sigma$ \citep{Sing2019}. Therefore, we use Equation \ref{eqn:mvolc1} to determine what value of $f_{\rm esc,\,i}\dot M_{\rm volc,\,i}$ would produce an absorption signal with $f_{\rm abs,\,i}=10\%$ as measured across a discrete, binned wavelength band to emulate real observations. For wavelengths $<300$~nm, we compute absorption across bands of width $0.1$~\AA, comparable to $\sim5$ resolution elements in COS grisms and STIS Echelle gratings \citep{STIS_book,COS_book}. At wavelengths $>300$~nm, we use bandwidths of $2$~nm, comparable to $\sim2$ resolution elements for WFC3/UVIS, WFC3/IR, and STIS/IR grisms \citep{STIS_book,2024wfci.book...16M}. The spectral energy distributions of GKM dwarf stars peak in the visible and infrared, falling off in blue and UV wavelengths unless significant non-local thermodynamic equilibrium (non-LTE) emission is present. Therefore, we bin a larger number of resolution elements at wavelengths $<300$~nm to achieve sufficient signal-to-noise ratio for constraining volcanic outgassing to an order of magnitude. We calculate the minimum detectable outgassing rate for atomic and ionic species relevant to volcanism \citep[e.g. carbon, oxygen, sodium, sulfur, and potassium; see][]{Bridge1979Sci...204..987B, Broadfoot1979Sci...204..979B} and present these values in Table \ref{tab:results2}. To estimate $f_{\rm abs,\,i}$, we calculate the absorption spectrum of each species using \texttt{Cthulhu} \citep{Cthulhu2024JOSS....9.6894A}. Pressure and thermal broadening of absorption lines is calculated assuming a plasma pressure of $p=1$~nbar and $T=1000$~K, chosen based on pressures and temperatures modeled for Io's atmosphere \citep{Strobel1994}. The exceptionally low pressure of the rarefied plasma results in line profiles that are sharp compared to the absorption features typically encountered in stellar and planetary spectra. Plasma tori pressures and temperatures are uncertain; we discuss this topic further in Section \ref{sec:metalsISM}.

\setlength{\tabcolsep}{4.0pt}
\begin{table*}[t]
\begin{rotatetable*}
\begin{center}
\caption{Required outgassing rates to absorb 10\% of the nominal stellar flux at a specified wavelength for planets in Table \ref{tab:results}.}\label{tab:results2}
\begin{tabular}{ccccccccccc} 
\hline
\hline
Name & C-I & C-II & O-I & Na-I & S-I & S-II & S-III & K-I & Total & Quasi-Steady Mass \\
&[ton~s$^{-1}$]&[ton~s$^{-1}$]&[ton~s$^{-1}$]&[ton~s$^{-1}$]&[ton~s$^{-1}$]&[ton~s$^{-1}$]&[ton~s$^{-1}$]&[ton~s$^{-1}$]&[ton~s$^{-1}$]&[Mt] \\
\hline
GJ 367 b & 0.7 & 0.7 & 2.6 & 1.4 & 0.9 & 16 & 26 & 1.9 & 51 & 715 \\ 
TOI-238 b & 17 & 17 & 64 & 34 & 22 & 396 & 639 & 46 & 1235 & 2898 \\ 
GJ 9827 b & 15 & 15 & 56 & 30 & 19 & 352 & 569 & 41 & 1098 & 2084 \\ 
LHS 1678 b & 6.3 & 6.5 & 24 & 13 & 8.1 & 150 & 241 & 18 & 466 & 732 \\ 
L 168-9 b & 22 & 23 & 86 & 46 & 29 & 535 & 864 & 63 & 1668 & 2275 \\ 
K2-36 b & 31 & 32 & 117 & 63 & 40 & 730 & 1178 & 86 & 2275 & 3035 \\ 
HD 23472 d & 70 & 72 & 266 & 143 & 90 & 1657 & 2675 & 194 & 5166 & 5142 \\ 
GJ 486 b & 18 & 18 & 67 & 36 & 23 & 418 & 674 & 49 & 1301 & 936 \\ 
HD 260655 b & 55 & 57 & 211 & 114 & 71 & 1316 & 2124 & 154 & 4102 & 2154 \\ 
Kepler-102 b & 177 & 183 & 674 & 363 & 228 & 4206 & 6788 & 493 & 13112 & 6553 \\ 
GJ 9827 c & 122 & 127 & 465 & 251 & 158 & 2902 & 4685 & 340 & 9049 & 3938 \\ 
Kepler-102 c & 311 & 323 & 1188 & 640 & 402 & 7412 & 11964 & 869 & 11147 & 7838 \\ 
HD 63433 d & 445 & 462 & 1697 & 914 & 575 & 10585 & 17086 & 1242 & 5333 & 7790 \\
\hline
\multicolumn{11}{p{1.1\textwidth}}{\small{\textbf{Notes. }The rates shown assume $f_{\rm esc}=1$ and are estimated for the following absorption lines: C-I: 166~nm. C-II: 134~nm. O-I: 130~nm. Na-I: 589~nm. S-I: 181~nm. S-II: 126~nm. S-III: 119~nm. K-I: 767~nm. Each line was selected for having the highest available cross section on the range 100--1800~nm where HST is sensitive. Only species which have absorption lines of sufficiently high cross-section (e.g. $\sigma\geq10^{-16}$~cm$^2$) to be detected with reasonable outgassing rates are included. Absorption cross sections $\sigma$ were computed using \texttt{Cthulhu} \citep{Cthulhu2024JOSS....9.6894A} and binned down to spectral resolution comparable to HST COS, STIS, and WFC3 instruments. For wavelengths shorter than $300$~nm, we bin every $0.1$~\AA\,to emulate HST COS and STIS Echelle observations. At wavelengths longer than 300~nm, we bin every 2~nm to emulate HST STIS grating and WFC3 grism observations. The linelists supplied to \texttt{Cthulhu} were sourced from VALD-3 \citep{VALD}, which relied on data from \cite{INTP}, \cite{MSb}, \cite{NIST10}, \cite{K04},
\cite{K12}, \cite{BPM}, \cite{GUES}, \cite{KP}, 
\cite{LN}, \cite{WSM}, \cite{LWa}, \cite{BQZ}, and \cite{ZCBS}. We also present the total mass loss rate summed across all species as well as the maximum expected torus quasi-steady mass. For reference the total mass loss rate observed from Io is $f_{\rm esc}\dot M_{\rm volc}\sim1$~ton~s$^{-1}$ \citep{Dessler1980Icar...44..291D} and the torus mass is $\sim2$~Mt \citep{BagenalDISTRO1980GeoRL...7...41B}.}} \end{tabular}
\end{center}
\end{rotatetable*}
\end{table*}

Volcanic outgassing rates spanning $\lesssim1$ to $\gtrsim10^4$~ton~s$^{-1}$ are predicted by Equation \ref{eqn:mvolc1} in Table \ref{tab:results2}, assuming complete injection with $f_{\rm esc}\simeq1$. For comparison, outgassing on Io typically injects $\sim1$~ton~s$^{-1}$ \citep{Dessler1980Icar...44..291D} of mostly sulfur and oxygen \citep{Bridge1979Sci...204..987B,Broadfoot1979Sci...204..979B} into the Jovian magnetosphere, while volcanism on Earth annually releases among other things 0.18--0.44~Gton of CO$_2$ \citep{Marty1998ChGeo.145..233M}, or equivalently $\sim$6--14~ton~s$^{-1}$. We note that comparable mass injection rates could produce detectable plasma tori around some of our identified targets. An Io-like rate of $\sim1$~ton~s$^{-1}$ of carbon and sodium is sufficient to create detectable C-I, C-II, and Na-I absorption for GJ~367~b, while an Earth-like production rate of $\sim$6--8~ton~s$^{-1}$ creates detectable C-I, C-II, and Na-I absorption for LHS~1678~b. Slightly higher rates of 10--20~ton~s$^{-1}$ would produce detectable S-II absorption for GJ~367~b and detectable C-I/C-II absorption for TOI-238~b, GJ~9827~b, and GJ~486~b. Species with weaker cross-sections, such as S-II and S-III, may be detectable at a modest mass injection rate of 26~ton~s$^{-1}$ for GJ~367~b but could only be detected around our other targets at more intense volcanic outgassing rates exceeding 100~ton~s$^{-1}$. For S-II and S-III, most targets would require very high rates $\gtrsim1000$~ton~s$^{-1}$ to produce detectable plasma tori. We do not anticipate these detections being made as such outgassing rates far exceed anything presently seen in the Solar System. Overall, Table \ref{tab:results2} suggests that Io- and Earth-like outgassing rates of 1--10~ton~s$^{-1}$ would be sufficient to construct detectable plasma tori around GJ~367~b and LHS~1678~b. Slightly more rigorous outgassing of 10--100~ton~s$^{-1}$ can create features around most of our targets, while the planets Kepler-102~b/c, GJ~9827~c, and HD~63433~d will need exceptional outgassing at rates $>>100$~ton~s$^{-1}$ to produce detectable circumstellar plasma tori. We note that as all values in Table \ref{tab:results2} were produced with Equation \ref{eqn:mvolc1}, the estimated outgassing rates for the neutral species should be understood to be uncertain, while more confidence may be placed in the rates for ionized species.

From Table \ref{tab:results}, the target exoplanet with the highest magnetospheric convection timescale $\tau$ and radiation magnetic confinement parameter $\zeta$ is GJ~367~b. Table \ref{tab:results2} shows that GJ~367~b has the lowest minimum volcanic outgassing rate required to be detectable by HST. To demonstrate the feasibility of our proposed exoplanet volcanic outgassing detection method, we examine this target in more detail next.

\subsection{GJ 367 b}
GJ~367~b is a sub-Earth planet with a radius of $0.699\pm0.024$R$_\oplus$ and a mass of $0.633\pm0.050$M$_\oplus$. This combination suggests that it is a ``super-Mercury'' with a massive iron core of radius fraction $86\pm5$\% \citep{Lam2021Sci...374.1271L,Goffo2023ApJ...955L...3G}. James Webb Space Telescope (JWST) MIRI/LRS eclipse data of this planet revealed a lack of heat redistribution implying a surface pressure upper bound of $p_{\rm surf}\lesssim0.5$~bar \citep{Zhang2024ApJ...961L..44Z}. The lack of a substantial atmosphere could be fully attributed to intense stellar X-ray and UV flux driving atmospheric escape on Gyr timescales \citep{Zahnle2017ApJ...843..122Z}, without requiring a low volatile content at formation \citep{Zhang2024ApJ...961L..44Z}. These observations are consistent with a scenario in which GJ~367~b is either bereft of an atmosphere or bears a tenuous atmosphere with low surface pressure. The airlessness or near-airlessness of GJ~367~b also does not contradict the possibility that it is constantly injecting mass into a plasma torus through volcanic outgassing of volatiles stored within the mantle.

GJ~367~b is on a short-period, eccentric orbit with $P_{\rm Orb}=0.32$~days and $e=0.060^{+0.070}_{-0.040}$ \citep{Goffo2023ApJ...955L...3G}. This orbit configuration could cause GJ~367~b to experience significant tidal heating at a rate $\dot E_{\rm heat}/\Im(\tilde k_2) \sim6\times10^{30}$~erg~s$^{-1}$, and therefore GJ~367~b may exhibit intense and widespread volcanism. Using our log-normal priors on $B_*$ and $\dot M_{\rm wind}$ as described in Section \ref{sec:confine}, we estimate the stellar Alfv\'en radius $R_A\simeq0.035^{+0.017}_{-0.012}$~AU. GJ~367~b orbits with a semimajor axis $a=0.00709\pm0.00027$~AU and is therefore within the Alfv\'en surface. Furthermore, we estimate $\zeta(a)\simeq4.43\times10^4$ and $\eta(a)\simeq7.91\times10^2$, sufficient to ensure magnetic confinement dominates over radiation and wind pressure. We conclude that GJ~367~b meets all of the criteria to construct and retain a circumstellar plasma torus.

\begin{figure*}\centering
  \includegraphics[width=01.0\linewidth]{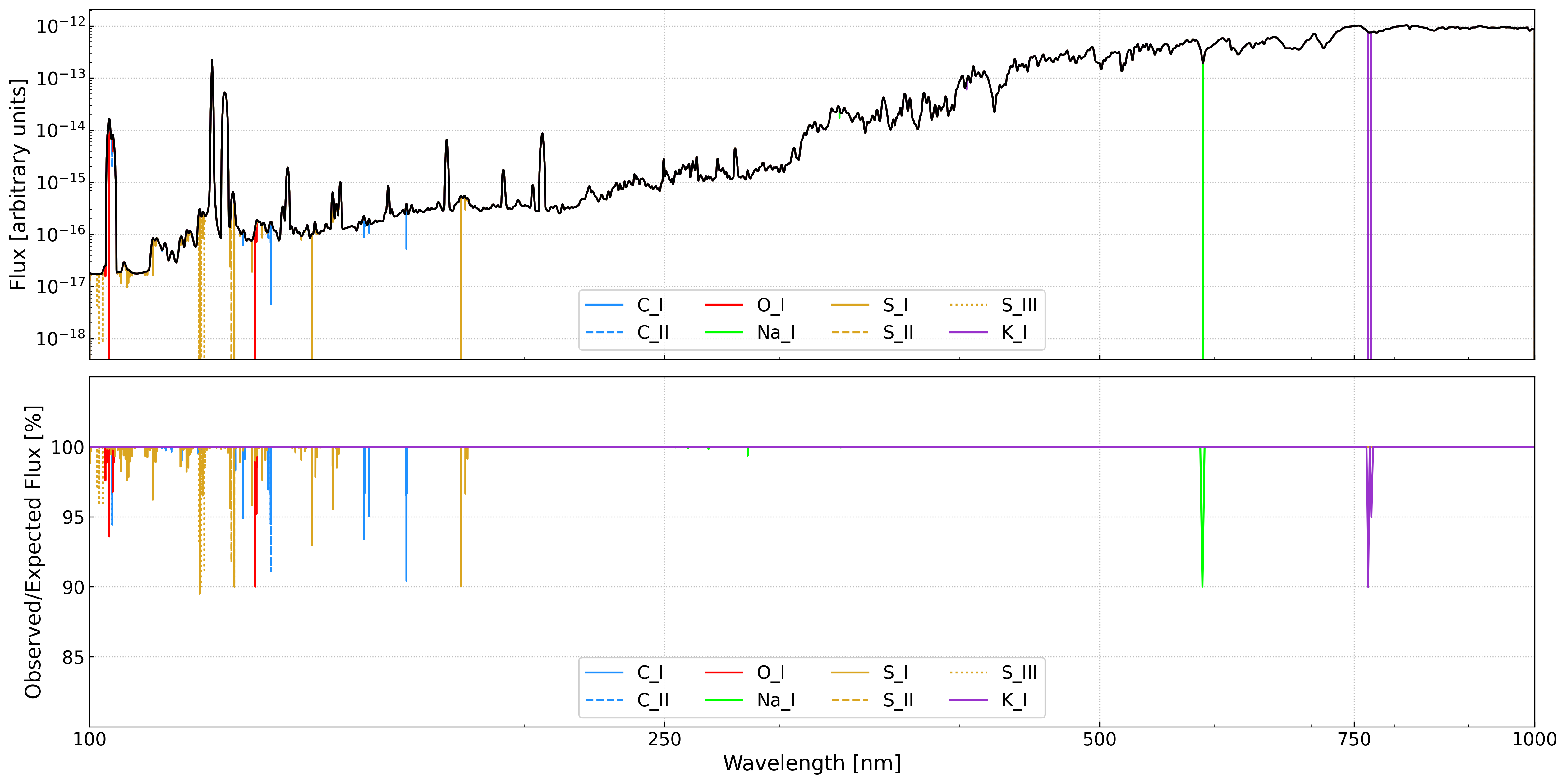}
  \caption{Measurable contamination by a hypothetical plasma torus in the GJ~367 system. Top: Model spectrum (S. Peacock, private communication) of the M dwarf star GJ~367~A (solid black), using stellar parameters $T_{\rm eff}=3560$~K, $\log{g}=4.78$ in cgs units, $M=0.44$~M$_{\odot}$, and $[$Fe/H$]=0$ which are comparable to those from \cite{Goffo2023ApJ...955L...3G}. Our spectrum was derived from a PHOENIX grid \citep{PHOENIX} model with updated treatment of M dwarf chromospheric non-LTE emission \citep[][]{PeacockFeb2019ApJ...871..235P,PeacockDec2019ApJ...886...77P}, with flux scaled to replicate GALEX FUV photometry of GJ~367~A. Colored lines denote spectral contamination by a circumstellar torus of plasma supplied by volcanism from GJ~367~b. The atomic species carbon (blue), oxygen (red), sulfur (yellow), sodium (green), and potassium (purple) are present in ionization states I (solid), II (dashed), and III (dotted) at the levels implied by the value of $\tau$ provided in Table \ref{tab:results} and the outgassing rates provided in Table \ref{tab:results2}. While some intrinsic stellar absorption features are expected at GJ~367~A's metallicity of $Z=-0.01\pm0.12$, a plasma torus would enhance the strength of select features based on its composition. Bottom: Absorption produced by a hypothetical GJ~367~b plasma torus relative to the expected flux as measured by HST COS, STIS, and WFC3 grisms and gratings. For wavelengths shorter than $300$~nm, we bin every $0.1$~\AA\,to emulate HST COS and STIS Echelle observations. At wavelengths longer than 300~nm, we bin every 2~nm to emulate HST STIS grating and WFC3 grism observations.}
  \label{fig:gj}
\end{figure*}

The magnetospheric convection timescale for material entering the magnetosphere at GJ~367~b's semimajor axis is $\tau\simeq162^{+194}_{-89}$~days (Table \ref{tab:results}). The calculations shown in Table \ref{tab:results2} suggest that a total mass injection rate $\simeq50$~ton~s$^{-1}$ is sufficient to produce absorption features with strength $f_{\rm abs}=10\%$ for absorption lines corresponding to sulfur, carbon, oxygen, sodium, and/or potassium. The steady state torus mass suggested by the timescale $\tau\simeq162^{+194}_{-89}$~days and injection rate $f_{\rm esc}\dot M_{\rm volc}\simeq50$~ton~s$^{-1}$ is approximately 700~Mt. The intense X-ray environment near GJ~367~A \citep{Poppenhaeger2024AA...689A.188P} provides a strong driver of atmospheric escape and ionization which will yield a high escape fraction $f_{\rm esc}$. Moreover, the JWST MIRI/LRS eclipse evidence \citep{Zhang2024ApJ...961L..44Z} supports the possibility of perfect or near-perfect atmospheric loss. We highlight these results in Figure \ref{fig:gj}, showing an example model of a torus-contaminated stellar spectrum for the GJ~367 system with the magnetospheric convection timescale and volcanic outgassing rates presented in Tables \ref{tab:results} and \ref{tab:results2}.

\section{Caveats and Further Work} \label{sec:caveats}
In this section we outline caveats to be considered for future searches of plasma tori. We also provide suggestions for further work that could be done to mitigate or remove some of these caveats and improve our understanding of these hypothetical structures.
\subsection{Verifying the Origin of Absorption Signatures}

The methodology outlined in this paper uses the strength of the detected carbon, oxygen, sodium, sulfur, and potassium features to infer the mass of a plasma torus and by extension the volcanic outgassing rate of the planet that produced it. However, this method hinges on two assumptions: that (i) all of the absorbing particles are components of the torus itself, and (ii) that the escaping material was recently produced by volcanism. In reality, there are two non-planetary sources of contamination that can also produce these absorption features and thereby imitate a plasma torus: stellar metallicity and the interstellar medium (ISM). Furthermore, we must consider conditions in which escape of a secondary atmosphere from a dormant planet could create a non-volcanic circumstellar plasma torus.

\subsubsection{Stellar Metallicity} \label{sec:metalsstars}
The majority of known stars are composed primarily of hydrogen, with a small amount of helium and trace amounts of metals. While these heavier metals typically concentrate in the stellar core and deep interior, photospheric metals are sufficiently abundant to be detectable spectroscopically. All species identified in this study as tracers of torus contamination -- the volcanic elements carbon, oxygen, sodium, sulfur, and potassium -- have also been identified in stellar spectra at metallicities comparable to solar. These features can manifest as absorption lines in the photosphere or as emission lines from excited chromospheric particles. Therefore, a degeneracy exists between the spectrum of a torus-contaminated star and that of a highly metallic star.

We consider population studies of stellar abundances as a means of resolving this degeneracy. Thousands of stellar abundance ratios for specific elements, including the five tracer species identified above, have been compiled in the Hypatia Catalog Database \citep{Hinkel2014AJ....148...54H}. These data reveal correlations between the abundance ratios of the volcanic elements and the iron abundance ratio [Fe/H], which is often taken as a proxy of overall stellar metallicity $Z$. A highly metallic star that is not contaminated by a torus should be expected to follow the same trends found in the populations compiled in the Hypatia Catalog Database. Conversely, torus contamination would cause selective enhancement of only the volcanic element absorption features. Therefore, a torus-contaminated star may be distinguishable from a metallic star by comparing its abundances against its [Fe/H] ratio. Contaminated stars should appear as outliers with abnormally high values of e.g. [S/H] or [Na/H] relative to their [Fe/H] value, while intrinsically metallic stars will likely follow population-level metallicity trends.

\subsubsection{Metals in the ISM} \label{sec:metalsISM}
Interstellar sodium absorption has been found to inflate the inferred sodium abundances in the atmospheres of hot Jupiters if not accounted for \citep[e.g.][]{Zhou2012MNRAS.426.2483Z, lange2022MNRAS.514.5192L}. Likewise, all other elements that we have identified in this study as tracers of extrasolar volcanism can also be encountered in the ISM. We therefore require a method to distinguish absorption signatures impressed on the stellar spectrum by a plasma torus from features created by the ISM. Spectroscopy with sufficiently high resolution to resolve relevant line profiles could break this degeneracy. The width of these atomic absorption features sensitively depends on the ambient pressure and temperature. ISM absorption lines are extremely narrow because the ISM itself is diffuse. Plasma tori are also diffuse but less so than the ISM. It is therefore likely that their absorption features would be broader than those of the ISM. If this is the case, then observations of a stellar spectrum contaminated by both the ISM and a plasma torus would yield two-tiered absorption features, with a narrow feature from the ISM centered in the trough of a broader feature from the plasma torus. The distinct profiles allow the ISM contribution to be estimated and removed so that the plasma absorption is isolated, comparable to the methods used in \cite[e.g.][]{Zhou2012MNRAS.426.2483Z, lange2022MNRAS.514.5192L}.

Unfortunately, the exact temperatures and pressures of the plasma in a circumstellar torus are uncertain. Therefore, it cannot be known for certain that plasma torus absorption features will be measurably broader than those of the ISM. If plasma torus absorption lines are not significantly broadened compared to ISM lines, then HST observations may not have sufficient spectral resolution to distinguish the two profiles from each other. 

Alternatively, one could obtain spectra of stars neighboring the target star, such as a resolved binary companion or a near-field star. Assuming that the ISM does not vary rapidly between the lines of sight of target and comparison stars, these comparison spectra could be used to measure local ISM contamination. If similar absorption features appear in the spectra of both the target star and the comparison stars, the source of the absorption is most likely the ISM. Conversely, if the target star shows selective enhancement of certain absorption features (e.g. sulfur or sodium) that do not appear in the comparison stars, the extant to which these features are enhanced could be used to infer the mass of a contaminating plasma torus.

We note that all the targets presented in Table \ref{tab:results} are located at distances $d\lesssim100$~pc from the Earth and therefore within the Local Bubble \citep{Lallement2003}. Metal abundances within the Local Bubble are known to be depleted to $n_{Z}\lesssim10^{-9}$~cm$^{-3}$ \citep{Welsh2010,draine2011}. Therefore, ISM contamination will be of order $\lesssim1\%$, considerably less significant than the $\sim10\%$ signals we are searching for. ISM contamination may cause slight overestimates of $\dot M_{\rm volc}$. However, we do not expect the ISM to create a false positive torus for any of the target exoplanets identified here nor dramatically skew the results of our interpretation.

\subsubsection{Escaping Secondary Atmospheres from Dormant Planets} \label{sec:lateescape}
Our method assumes that the source of the torus material is recently-outgassed material from the mantle. Therefore, the torus mass injection rate can be directly linked to the volcanic outgassing rate. However, it is possible that the planet built a massive secondary atmosphere via previous, now-dormant volcanic activity. Photoevaporation of this secondary atmosphere would strongly resemble escape from ongoing volcanism. Secondary atmosphere composition is at least partially set by volcanism and so some insight into the interior composition is still possible in this scenario. However, there will be highly uncertain contributions from other processes that modify secondary atmosphere composition such as photochemistry, diminishing our ability to directly connect the compositions of the mantle and torus.

In the absence of replenishment by volcanism, plasma tori sustained by photoevaporation of a secondary atmosphere are expected to be massive but short-lived. All targets in Table \ref{tab:results} orbit within the stellar magnetosphere at $a/R_*\leq20$, and are therefore subject to intense stellar radiation which could drive atmosphere escape at rates far surpassing the needed volcanic outgassing rates (see Appendix \ref{sec:atmosescape} for further discussion). For example, the XUV flux acting on GJ~367~b is expected to drive atmospheric mass loss of $\sim1000$~ton~s$^{-1}$ \citep{Poppenhaeger2024AA...689A.188P}, $\sim20$ times greater than the total volcanic outgassing rate needed to create a detectable torus. At this mass injection rate, Equations \ref{eqn:torusmass} and \ref{eqn:mvolc1} predicted a torus quasi-steady mass $\gtrsim10$~Gt, much larger than the quasi-steady mass of $\sim700$~Mt shown in Table \ref{tab:results2}. Such a massive torus would necessarily be short-lived, as this photoevaporation rate could strip even 100 Venusian atmospheres' worth of mass in under 1~Gyr. Apart from HD~63443~d which is thought to be under 550~Myr in age \citep{Capistrant2024AJ....167...54C}, the targets present in Table \ref{tab:results2} are expected to be over 1~Gyr in age. Therefore, it is unlikely that any target apart from HD~63443~d could still be maintaining a torus through photoevaporation alone.

In contrast, tori sustained by active volcanism instead of atmospheric photoevaporation are expected to be much longer-lived with lower quasi-steady masses. Sequestration beneath the mantle prevents volatiles from being rapidly stripped by photoevaporation. The rate at which mantle volatiles are lost is instead controlled by volcanic processes. It is estimated that Earth's mantle contains $10^{22}$--$10^{23}$~kg of sequestered volatiles \citep[e.g.][]{ZhangZindler1993EPSL.117..331Z,Sleep2001JGR...106.1373S,Marty2012EPSL.313...56M,ZHANG201437}, equivalent to over 1000 times the mass of Earth's present atmosphere. Suppose that GJ~367~b possesses a similarly massive volatile reservoir in its mantle. Then volcanic outgassing at the detectable rate of $\sim$10--100~ton~s$^{-1}$ (Table \ref{tab:results2}) could sustain a plasma torus of quasi-steady mass 200~Mt$\lesssim M_{\rm torus}\lesssim2$~Gt for $\geq3$~Gyr. In the case of the young exoplanet HD~63433~d, a plasma torus detection could not confirm whether the planet was volcanically active or not. Active volcanism is required to justify plasma tori detection in all of our older targets in Table \ref{tab:results}, as the intense stellar radiation environment will have depleted any secondary atmosphere that could have provided mass otherwise.

\subsection{Variety of Composition for Outgassed Material}
We have focused our discussion on five elements known to be major products of volcanism on Earth and Io: carbon, oxygen, sodium, sulfur, and potassium. In practice, any element that can be sequestered in the mantle should be considered a candidate for volcanic outgassing. For example, hydrogen in the form of water vapor is also a significant component of eruptions on Earth; we chose not to consider it here since hydrogen escape is a less ambiguous sign of volcanism and can easily be confused with evaporation of a primordial atmosphere \citep[e.g.,][]{Bourrier2018...620A.147B}. A relatively smaller amount of nitrogen and other noble gasses are also released by Earth's volcanism, and dust ejecta produce silicate grains as well. Future studies could incorporate these cross sections into the analysis described in Sections \ref{sec:analysis} and \ref{sec:results}.

Additionally, not all elements highlighted here are necessary for proof of volcanic activity. Earth's volcanoes do not produce significant quantities of sodium or potassium, while Io's volcanoes and torus show no carbon despite the fact that observations to date would have detected it if present \citep{Kesz2023ASSL..468..211K}. Io also shows a lack of hydrogen and nitrogen \citep{Kesz2023ASSL..468..211K}, indicating an apparent depletion of these volatiles within Io's interior. Likewise, a mixture of element detections and nondetections in a circumstellar plasma torus reveals both the abundant and depleted elements within the planet's interior.

\subsection{Simplifications to Magnetospheric Convection Theory}
We noted in Section \ref{sec:timescale} that the \cite{Hill1981} magnetospheric convection timescale was an underestimate that did not account for mechanisms which delay magnetospheric convection. Suppose a plasma torus is observed and the estimated magnetospheric convection timescale $\tau$ is used to infer the outgassing rate $\dot M_{\rm volc}$ with Equation \ref{eqn:mvolc1}. If $\tau$ is larger than estimated, the derived outgassing rate will be an overestimate. Improved methods of estimating $\tau$ \citep[e.g., the radial Fokker-Planck formulation of][]{Schreier1998JGR...10319901S} could be applied to reduce overestimation. The \cite{Hill1981} method underestimates $\tau$ for Io by a factor of $\sim100$. If we assume that the timescales in Table \ref{tab:results} are similarly underestimated, then outgassing rates are overestimated by no more than a factor of 100.

We also simplified the discussion of the transport by assuming that the torus material distributes across an idealized torus geometry with uniform number density $n$. The Io plasma torus is known to exhibit significant radial and vertical variations in number density \citep{Shemansky1980ApJ...236.1043S,Strobel1980ApJ...238L..49S}. These radial and vertical variations arise from inhibited radial magnetospheric convection \citep{Thomas2004jpsm.book..561T} and vertical diffusion along magnetic field lines \citep{BagenalDISTRO1980GeoRL...7...41B}, respectively. This simplified transport treatment affects how the measured column density is interpreted and thereby affects the inference of the torus quasi-steady state mass. While alternative models of number density distribution could be applied, the derived quasi-steady state mass estimate generally changes by less than an order of magnitude and is therefore contained within our uncertainty.

\subsection{Unconstrained Atmospheric Escape Fraction}
Our methodology measures the torus mass injection rate $f_{\rm esc}\dot M_{\rm volc}$. Without measuring the atmospheric escape fraction $f_{\rm esc}$, we cannot know the exoplanet's total volcanic outgassing $\dot M_{\rm volc}$. We established in Section \ref{sec:methods} that photevaporation is likely to drive significant atmosphere loss and thus $f_{\rm esc}\simeq1$ is the most likely possibility; nevertheless, $f_{\rm esc}<1$ is a possibility we cannot fully disprove. While the technique proposed here cannot measure $f_{\rm esc}$, we can place a lower limit on $f_{\rm esc}$. Smaller $f_{\rm esc}$ values imply higher values of $\dot M_{\rm volc}$. A terrestrial-mass planet is unlikely to outgas a significant fraction of its own mass over its lifetime. Therefore, $f_{\rm esc}$ small enough to imply a cumulative outgassed mass greater than $10\%$ of the planet's total mass should not be considered. Further work on this topic could constrain the escape fraction, e.g. by using models of photoevaporation \citep[e.g.,][]{Bourrier2013AA...557A.124B,WangDai2018ApJ...860..175W} or by employing MHD simulations \citep{Athenaplusplus2020ApJS..249....4S} to estimate the plasma heating rate.

\subsection{Torus Geometry Uncertainties}
We have assumed that circumstellar plasma tori are perfectly toroidal in form, with no spatial or temporal variations. This includes equating the plasma entry distance to the planet semimajor axis $a$. Variations in entry distance would arise in an eccentric orbit. For an order of magnitude estimate of torus mass, these variations can be neglected if $e\lesssim0.10$. We have also simplified the Alfv\'en surface to a sphere with radius $R_A$. However, the Alfv\'en surface of the Sun is spatially and temporally irregular \citep[e.g.,][]{Goelzer2014JGRA..119..115G, Deforest2014ApJ...787..124D, Liu2021ApJ...908L..41L, Cranmer2023SoPh..298..126C}. The Alfv\'en surfaces of other stars are also predicted by models to be irregular \citep[e.g.,][]{vidotto2014MNRAS.438.1162V}.

In reality the geometry of the Io plasma torus is highly complex. In addition to the warm torus, there is a cold inner torus and narrow ``ribbon'' \citep{BagenalDols2020}. Azimuthal and temporal variability are both requirements and consequences of magnetospheric convection \citep{Hill1981}. As described in Section \ref{sec:supply}, torus plasma is not supplied directly from Io's plumes, but by ionization of Io's neutral cloud of molecules extending a few times its radius ahead and behind its orbit \citep[see Section 4.1 of][for details]{BagenalDols2020}. An analogous neutral cloud in circumstellar plasma tori may produce detectable molecular absorption signatures. These features would appear in the IR and therefore be amenable to characterization with JWST.

Section \ref{sec:analysis} described a series of measurements that could detect and preliminarily characterize a plasma torus, including (i) measuring phase-resolved neutral cloud absorption by atomic species and (ii) taking ``snapshot'' measurements of torus atomic and ionic absorption. These UV-optical observations are sufficient to confirm the existence of the torus and obtain initial estimates of its steady state mass and mass injection rate. Characterization of the spatial and temporal irregularities of the torus could then be carried out in more detail through the following observations:
\begin{itemize}
    \item \textbf{IR transit ingress and egress of source exoplanet:} This follow-up observation would be capable of detecting molecules in the neutral cloud if present. Transits viewed in 4.0--4.5$\mu$m and/or 7.5--8.0$\mu$m would have asymmetric or extended ingresses and egresses as a result of molecular absorption if a neutral cloud of CO$_2$ and/or SO$_2$ were present. A JWST MIRI/LRS or NIRSpec observation could readily capture this phenomenon. Alternatively, the absence of neutral cloud molecules would suggest that photodissociation operates rapidly in circumstellar tori relative to the atmospheric material diffusion timescale.
    \item \textbf{IR transmission spectrum of source exoplanet's atmosphere:} This observation would constrain the composition and surface pressure of an atmosphere if present. This would provide constraints on photoevaporation and plasma heating. It would also resolve ambiguity in $f_{\rm esc}$; the lack of an atmosphere would imply that photoevaporation and/or plasma heating operated uninhibited to achieve $f_{\rm esc}\rightarrow1$ as expected. In that case the measured mass injection rate would be equal to the planet's total volcanic activity.
    \item \textbf{Time-resolved UV and visible spectroscopy of the out-of-transit stellar spectrum:} Measuring the torus-contaminated stellar spectrum as a function of magnetosphere rotation phase could reveal variations in the strength of atomic absorption features. This would produce azimuthally-resolved maps of torus column density. These azimuthal variations could correlate with (i) the planet's orbit and/or (ii) spatial and temporal irregularity in the Alfv\'en surface and magnetic activity of the star. Therefore, they could provide an exquisite probe into the structure of the stellar magnetosphere. This work would primarily be accessible to HST COS and STIS which are sensitive to the 100- to 200-nm range, but the visible lines of Na (589~nm) and K (767~nm) would be accessible both by HST and by ground-based high-resolution spectroscopy. The long rotation periods of our targets may prohibit fully phase-resolved maps from being acquired, but partially-resolved maps can still reveal stellar activity through e.g. uncovering flare-opened gaps or flare-damaged torus sections.
\end{itemize}

\subsection{Feasibility of Proposed Observations for Current and Future Observatories}
In this paper we demonstrated that the proposed research is astrophysically possible; known exoplanets outgassing at realistic rates into their host stars' magnetospheres would construct circumstellar plasma tori of sufficient column density to produce absorption at the level of $\gtrsim10\%$ of the UV pseudo-continuum. However, this method relies on recovery of the stellar UV pseudo-continuum flux in order to measure the torus-imprinted absorption signatures. The target M stars identified in Table \ref{tab:results} are similar in spectral class, distance, and V magnitude to the stars targeted by the MUSCLES and Mega-MUSCLES Treasury Surveys \citep{MUSCLES2016ApJ...820...89F}. These surveys used HST COS and STIS to measure the UV spectra of multiple M dwarf stars, and found that HST was not able to recover 100--200 nm continuum flux for several targets due to encountering HST instrument noise floors \citep{Wilson2025ApJ...978...85W}. We therefore expect HST COS and STIS to encounter challenges in recovering the UV pseudo-continuum fluxes of our target stars.

However, while the UV pseudo-continuum flux predicted by standard PHOENIX grid models \citep[e.g.,][]{PHOENIX1999,PHOENIX,PHOENIX2013} and empirical MUSCLES/Mega-MUSCLES models \citep[e.g.,][]{MUSCLES2016ApJ...820...89F, MUSLCES2016_2, MUSCLES2016_3, Wilson2025ApJ...978...85W} falls below the HST STIS/NUV-MAMA noise floor, our model spectrum for GJ~367~A shown in Figure \ref{fig:gj} has sufficient UV pseudo-continuum flux to be above the noise floor. We recommend that follow-up work carefully consider the UV spectra of these stars using a variety of M dwarf stellar models \citep[e.g.,][]{PeacockFeb2019ApJ...871..235P,PeacockDec2019ApJ...886...77P, Hintz2023} to properly assess which may have recoverable UV pseudo-continuum fluxes. For those that do fall below the COS and STIS noise floors, tori absorption in the UV will be undetectable until the advent of new and more sensitive instrumentation. As a result, tori searches for these targets may be confined to optical absorption by neutral alkali species.

In the absence of optical absorbers, torus \textit{emission} rather than absorption may be the preferred avenue for studying these targets. HST STIS/FUV-MAMA low-resolution gratings can recover near-UV emission lines with line-integrated flux $\geq10^{-16}$~erg~cm$^{-2}$~s$^{-1}$ at an SNR of 10 in under 10 hours of exposure, according to the Exposure Time Calculator. Torus emission lines could outshine unrecovered UV pseudo-continuum flux for dim M dwarf targets, making interpretation of the line much cleaner. Thus, the lack of stellar UV pseudo-continuum becomes a perk for emission surveys rather than a limitation. Future work should investigate the conditions under which circumstellar plasma torus emission reaches a flux level detectable by HST and, following the methodology presented here, develop an analogous equation to Equation \ref{eqn:mvolc1} that connects detected non-LTE emission intensity to volcanic outgassing rate.

Absorption studies may become more feasible for UV-dim targets in the coming decades. In particular, proposed UV-optical-IR observatories such as the Habitable Worlds Observatory (HWO) are recommended to be sensitive to UV flux at 1000--2000~\AA \citep{LUVOIR,HWO2024} down to $\simeq10^{-18}$~erg~cm$^{-2}$~s$^{-1}$ at medium spectral resolution \citep{LUVOIR2}. Near-future UV observatories, such as the planned Ultraviolet Explorer (UVEX) \citep{UVEX} and proposed concepts like UV-SCOPE \citep{UVSCOPE}, also aim to detect $\simeq10^{-18}$~erg~cm$^{-2}$~s$^{-1}$ at 1000-2000~\AA. Thus, torus UV absorption studies may find their footing in the early 2030s as instrumentation advances.

\section{Conclusions} \label{sec:conclusions}
We have presented a novel method of constraining exoplanet volcanic outgassing rates through detection and mass estimation of circumstellar plasma tori. This work generalizes the physical processes operating in the Jupiter-Io plasma torus to analogous structures in star-planet systems. This hypothetical torus would disperse on timescales much shorter than geologic timescales and therefore requires ongoing and active volcanism to maintain. The detection of such a plasma torus would not only provide decisive evidence of active volcanism in the system but also offer order-of-magnitude constraints on the outgassing rate of the source exoplanet.

We presented candidate exoplanets with properties amenable for constructing plasma tori (Table \ref{tab:results}). These high-priority  exoplanets are (i) terrestrial, (ii) orbiting within the stellar Alfv\'en radius, and (iii) eccentric. Therefore, these exoplanets are susceptible to tidal heating that can sustain active volcanism over several Gyr, and the material they outgas is susceptible to magnetic confinement. The eccentric super-Mercury GJ~367~b appears to be one of the most compelling candidates for hosting a plasma torus built through tidal heating-driven volcanism. Moreover, we have shown that a physically-realistic volcanic outgassing rate of 1--10~ton~s$^{-1}$ could produce absorption features strong enough to be detected by multiple active observatories operating in the UV-optical, including HST and ground-based high-resolution spectrographs, provided that the stellar UV pseudo-continuum can be detected. Measurements of circumstellar plasma tori are not phase-constrained --- unlike most exoplanetary observations --- and can therefore be collected at any time. This aspect lends flexibility to scheduling such visits with highly oversubscribed space-based observatories.

We have also presented some of the limitations of this methodology, as well as opportunities for synergistic studies. The technique we use to estimate magnetospheric convection timescales $\tau$ could overestimate outgassing rates and therefore could be updated. Our simplified treatment of magnetospheric convection also employs uniform radial and vertical number density profiles which could be improved by a more nuanced treatment of transport to better interpret measured column densities. UV-optical observations that are phase-resolved with respect to stellar magnetosphere rotation phase could reveal azimuthal asymmetries if present, which in turn would provide insight into stellar activity and the structure of the magnetosphere. Transit observations in the IR could place constraints on the presence, structure, and composition of (i) an atmosphere and (ii) molecules within any structures analogous to Io's neutral cloud.

This paper highlights the pressing need for UV successors to HST. Many of the targets suitable for this work are dim at UV wavelengths and at risk of being undetectable by HST's sensitivity. With the launch of HWO still over a decade away, there exists demand for small- and medium-class UV missions to bridge the gap between the end of HST's operation and HWO's first light. UV-sensitive space telescopes can provide vital characterization of exoplanetary atmospheres and exospheres, crucial to understanding atmospheric aerosols \citep[][]{Ohno2020ApJ...895L..47O}, characterizing the effects of stellar contamination on planetary transmission spectra \citep[][]{Rackham2018ApJ...853..122R,Rackham2019AJ....157...96R}, and detecting the spectral signatures of circumstellar plasma tori as proof of extrasolar volcanism, as shown here.

This work lays the foundation for a new method of constraining the volcanic outgassing rates of extrasolar terrestrial planets. If successful, such constraints would enable estimates of the internal heating and composition of extant worlds beyond the Solar System for the first time.

\section*{Acknowledgements}

We thank John Noonan, Ryan MacDonald, Jeff Valenti, and Fred Adams for informative and insightful discussion that facilitated this work. We also thank Sarah Peacock for sharing an updated model of M dwarf ultraviolet flux which contributed greatly to the development of Figure 6. D.Z.S. is supported by an NSF Astronomy and Astrophysics Postdoctoral Fellowship under award AST-2303553. This research award is partially funded by a generous gift of Charles Simonyi to the NSF Division of Astronomical Sciences. The award is made in recognition of significant contributions to Rubin Observatory’s Legacy Survey of Space and Time. This work has made use of the VALD database, operated at Uppsala University, the Institute of Astronomy RAS in Moscow, and the University of Vienna. This research has made use of the NASA Exoplanet Archive, which is operated by the California Institute of Technology, under contract with the National Aeronautics and Space Administration under the Exoplanet Exploration Program. The authors thank the anonymous reviewer for their thorough and insightful comments that greatly improved the quality of this work.

All software used to reproduce the plots and results in this work can be found on Zenodo via\dataset[DOI: 10.5281/zenodo.15497840]{https://doi.org/10.5281/zenodo.15497840} and on GitHub\footnote{https://github.com/AbbyBoehm/BSL2025PlasmaTori}.

This research made use of observations from the \textit{Galaxy Evolution Explorer} (GALEX), obtained from the MAST data archive at the Space Telescope Science Institute, which is operated by the Association of Universities for Research in Astronomy, Inc., under NASA contract NAS 5–26555. The data used in this research can be found at\dataset[DOI: 10.17909/axwc-fw08]{https://doi.org/10.17909/axwc-fw08}.

\clearpage

\bibliography{sample631}{}
\bibliographystyle{aasjournal}

\appendix

\section{Driving atmospheric escape prior to the onset of plasma heating} \label{sec:atmosescape}
Following the formation of a plasma torus, atmospheric escape will be sustained by plasma heating \citep{Pospie1992GeoRL..19..949P}. However, the existence of a plasma torus is a prerequisite for plasma heating. Therefore, a different atmospheric escape mechanism must operate before plasma heating can begin. Here we provide an overview of two mechanisms that would initiate atmospheric escape: thermal escape and cumulative XUV-driven escape.

\subsection{Thermal Escape}
\begin{figure}[h!]
    \centering
    \includegraphics[width=1.00\columnwidth]{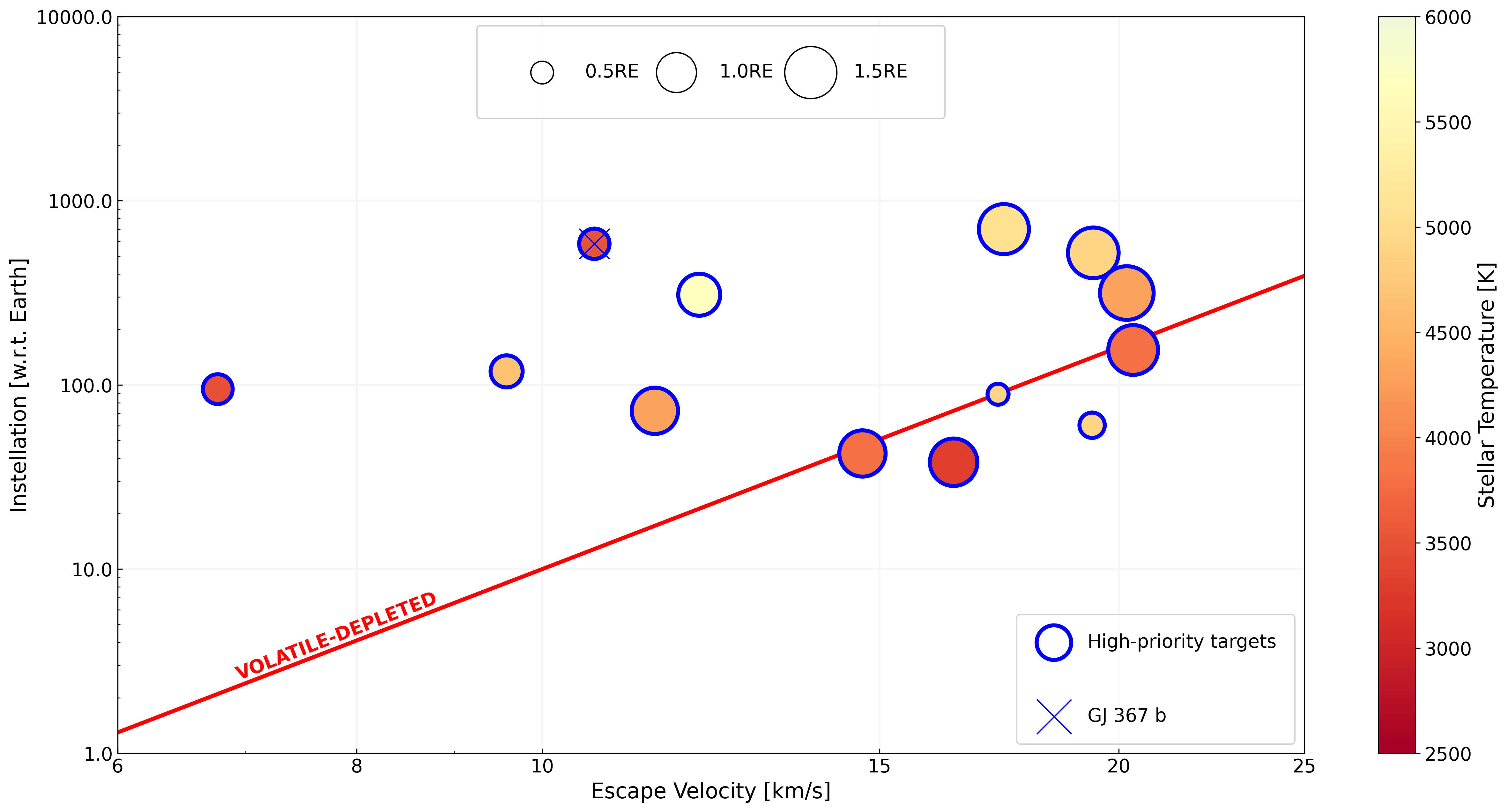}
    \caption{The 13 high-priority exoplanets highlighted in this study as they appear on the thermal cosmic shoreline. With increasing vertical distance above the shoreline, the likelihood of substantial volatile depletion rises.}
    \label{fig:thermal}
\end{figure}

Atmospheric escape occurs when the atmospheric  particles' thermal velocities, $v_{\rm th}$  exceed the escape velocity, $v_{\rm esc}$. The ``cosmic shoreline'' that separates volatile-rich worlds from volatile-depleted worlds is believed to follow an $I\propto v_{\rm esc}^4$ law with an unknown constant of proportionality. We estimate the shoreline following the methodology of  \cite{Zahnle2017ApJ...843..122Z} to be
\begin{equation} \label{eqn:shore}
    \frac{I}{I_{\oplus}} \simeq 10^{-3}\bigg(\frac{v}{1 \textrm{ km~s$^{-1}$}}\bigg)^4\,.
\end{equation}
In Equation \ref{eqn:shore}, $I_{\oplus}$ denotes the instellation present-day Earth receives. In Figure \ref{fig:thermal} we plot $I$ against $v_{\rm esc}$ for the exoplanets in Tables \ref{tab:results} and \ref{tab:results2}, marking the shoreline with Equation \ref{eqn:shore}.

Many of our target exoplanets lie above or near the thermal escape cosmic shoreline. It is therefore likely that thermal escape will be sufficient to deplete all of our high-priority targets of their atmospheric volatiles.

\subsection{XUV-Driven Escape}
\begin{figure}[h!]
    \centering
    \includegraphics[width=1.00\columnwidth]{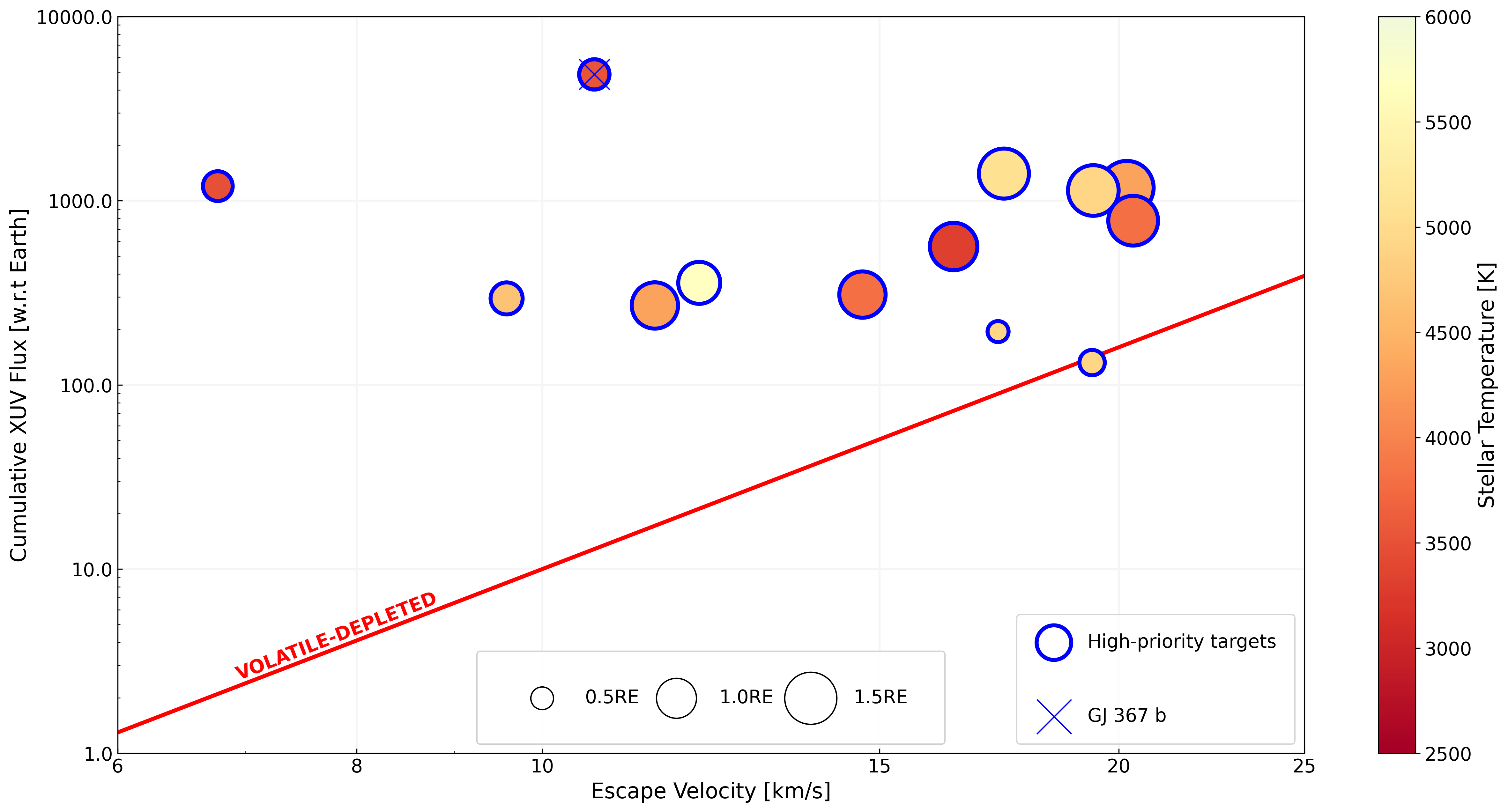}
    \caption{The 13 high-priority exoplanets highlighted in this study as they appear on the XUV-driven cosmic shoreline. With increasing vertical distance above the shoreline, the likelihood of substantial volatile depletion rises.}
    \label{fig:XUV}
\end{figure}

Stars undergo considerable temporal evolution both in their total luminosity and in their spectral energy distribution. Stars emit more intense X-ray and extreme UV (XUV) flux in earlier phases of their lifetime. Moreover, the duration of the  intense XUV emission phase depends on stellar mass \citep[see e.g.][for an overview of this issue]{Luger2015AsBio..15..119L}. \cite{Zahnle2017ApJ...843..122Z} calculated a cosmic shoreline which incorporates the high activity of pre-main sequence stars. This shoreline is similar to the thermal escape shoreline with the power law proportion $I_{\rm XUV}\propto v_{\rm esc}^4$. Equation 27 of \cite{Zahnle2017ApJ...843..122Z} presents a scaling relation between the stellar luminosity $L_*$ and present-day instellation $I$ for the cumulative XUV instellation:
\begin{equation} \label{eqn:xuv}
    I_{\rm XUV}=I\bigg(\frac{L_*}{L_\oplus}\bigg)^{-0.6}\,.
\end{equation}
In Figure \ref{fig:XUV} we show $I_{\rm XUV}$ against $v_{\rm esc}$ for the exoplanets in Tables \ref{tab:results} and \ref{tab:results2}, marking the shoreline with Equation \ref{eqn:shore} but substituting $I_{\rm XUV}$ in place of $I$.

All but one of our targets lie above the XUV-driven shoreline; the remaining target is on the shoreline. We therefore conclude that all 13 high-priority targets identified in Table \ref{tab:results} had sufficient atmospheric escape to initiate plasma heating and deplete their initial atmospheres within 1~Gyr of formation.

\section{Stellar magnetosphere priors and resultant Alfv\'en radius posteriors} \label{sec:alfvenposterior}
\begin{figure}[h!]
    \centering
    \includegraphics[width=0.3\linewidth]{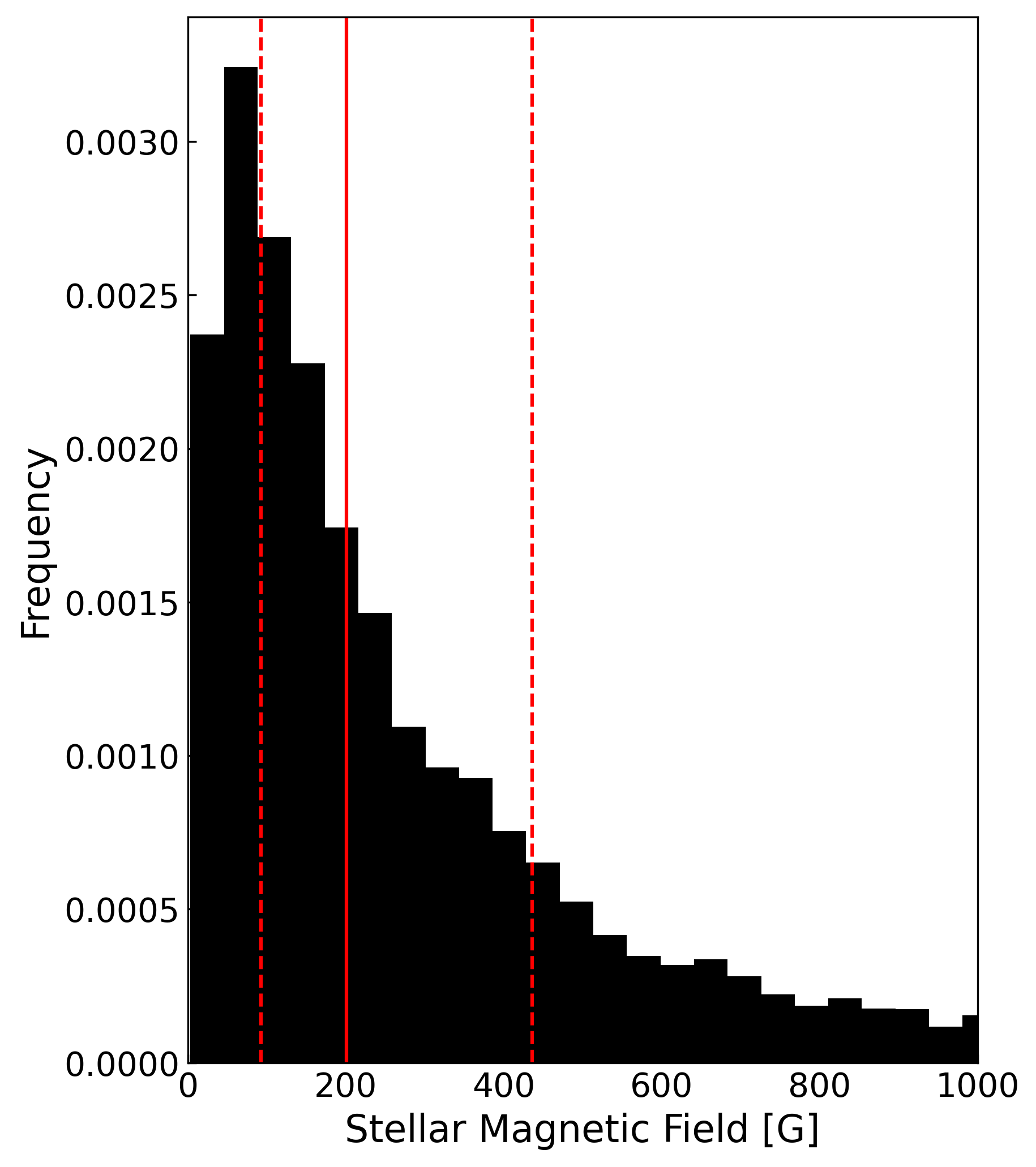}
    \includegraphics[width=0.276\linewidth]{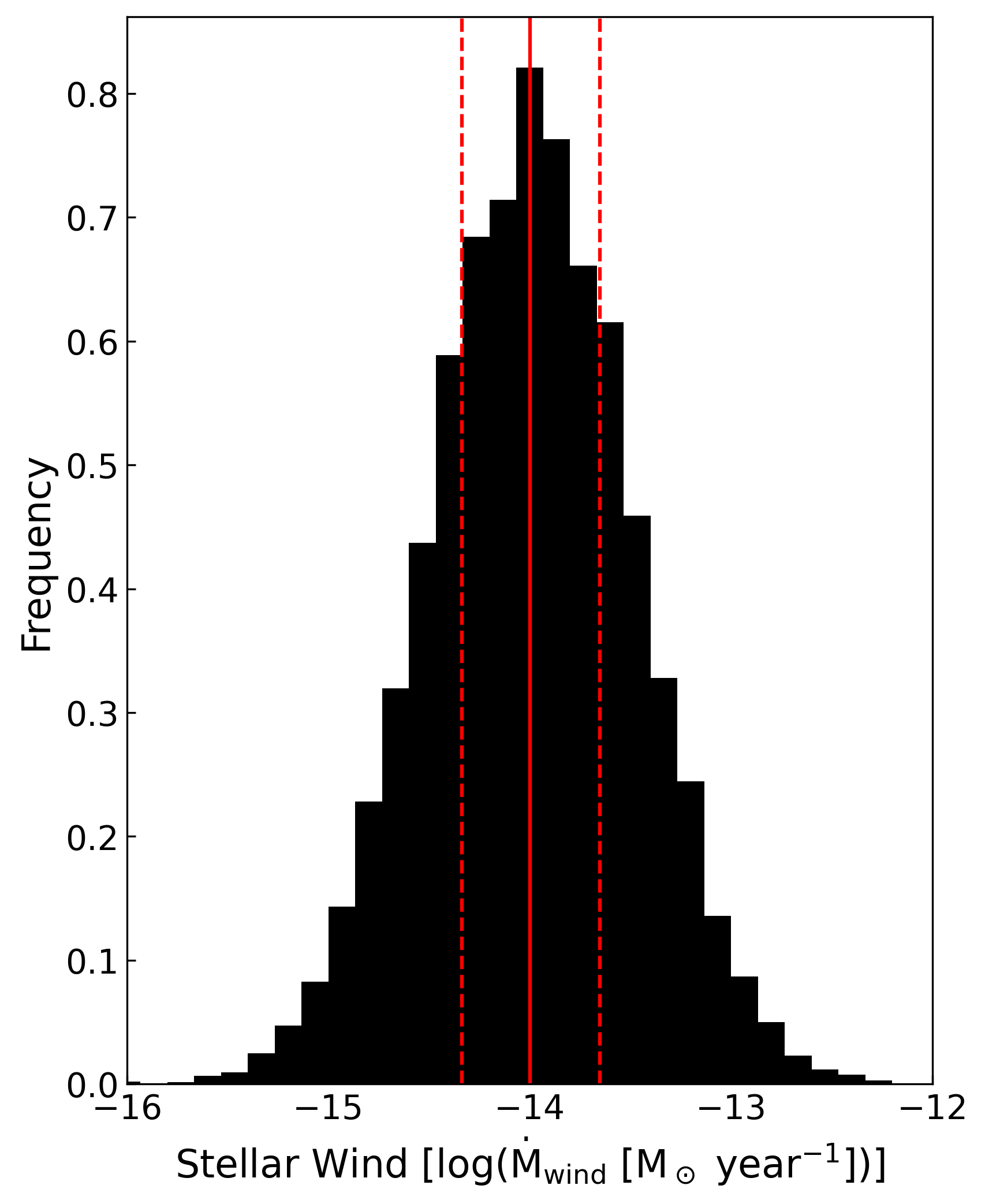}
    \includegraphics[width=0.28\linewidth]{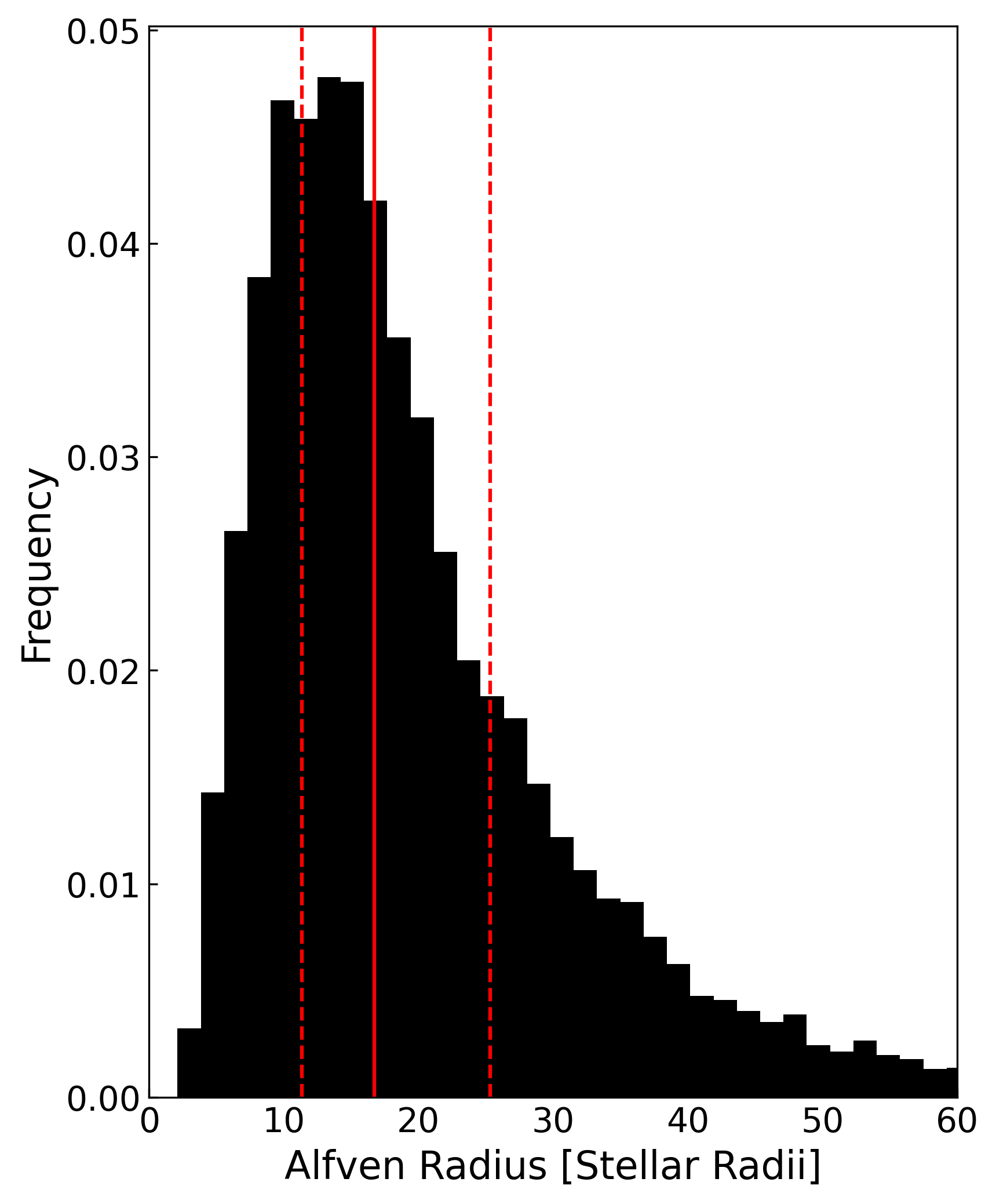}
    \caption{$B_*$ and $\dot M_{\rm wind}$ priors and resultant $R_A$ posterior for GJ~367~A \citep[$R_*=0.458$~R$_{\oplus}$, $v_{\rm \infty}\simeq616$~km~s$^{-1}$; from][]{Goffo2023ApJ...955L...3G}. Left: Log-normal prior for $B_*$. As an M dwarf star with $P_{\rm *}=51.3$~days, the most probable value of magnetic field strength for GJ~367~A is $B_*=200^{+200}_{-100}$~G \citep{Lehmann2024MNRAS.527.4330L}. Middle: Log-normal prior for $\log(\dot M_{\rm wind} [\textrm{M}_\odot~\textrm{yr}^{-1}])$. The measured x-ray flux of GJ~367~A \citep{Poppenhaeger2024AA...689A.188P} and likely $B_*\simeq200$~G suggests a most likely stellar wind mass loss rate $\log{(\dot M_{\rm wind} [\textrm{ M}_\odot~\textrm{yr}^{-1}])}=-14\pm0.3$ and disfavors loss rates greater than $10^{-13}\textrm{ M}_\odot~\textrm{yr}^{-1}$ \citep{Bloot2025arXiv250214701B}. Right: The resultant posterior distribution of the Alfv\'en radius. The expected Alfv\'en radius in stellar radii is $16.7^{+8.6}_{-5.3}$, equivalent to an orbit distance of about $0.035$~AU.}
    \label{fig:priorposterior}
\end{figure}

We sample the Alfv\'en radius by solving Equation \ref{eqn:AlfvenR} for various values of $\eta_*$ as computed by Equation \ref{eqn:etastar} to account for uncertainty in the properties of the stellar magnetosphere. Our $\eta_*$ posterior is constructed from log-normal priors on $B_*$ and $\dot M_{\rm wind}$. For $B_*$ we sample a prior with mean of $200$~G and a width of $\pm0.5$~dex, while our $\dot M_{\rm wind}$ prior has a mean $\log(\dot M_{\rm wind} [\textrm{M}_\odot~\textrm{yr}^{-1}])$ of $-14$ and a width of $\pm0.5$~dex. In Figure \ref{fig:priorposterior} we present the sampled $B_*$ and $\dot M_{\rm wind}$ priors and the $R_A$ posterior for GJ~367~A constructed from these priors via Equations \ref{eqn:etastar} and \ref{eqn:AlfvenR}.

\end{document}